\documentclass[pra,aps,superscriptaddress,nofootinbib,longbibliography,notitlepage,twocolumn]{revtex4-2}

\usepackage[T1]{fontenc}
\usepackage{lmodern}
\usepackage{amsmath,amssymb}
\usepackage{braket}
\usepackage{graphicx}
\usepackage[colorlinks,linkcolor=blue,citecolor=blue,urlcolor=blue]{hyperref}
\usepackage{orcidlink}

\def\ii{{\rm i}}  \def\ee{{\rm e}}
\def\Ree{{\rm Re}}  \def\Imm{{\rm Im}}
\def\xx{\hat{\bf x}}  \def\zz{\hat{\bf z}}
\def\kB{{k_{\rm B}}}  \def\EF{{E_{\rm F}}}
\def\Eb{{\bf E}}  \def\Jb{{\bf J}}  \def\mb{{\bf m}}  \def\pb{{\bf p}}  \def\rb{{\bf r}}
\def\ww{\omega}  \def\vep{\varepsilon}  \def\mydeg{^{\circ}}
\def\Hm{\mathcal{H}}
\def\phiind{\phi^{\rm ind}}  \def\phiext{\phi^{\rm ext}}  \def\rhoind{\rho^{\rm ind}}
\newcommand{\pd}[2]{\mathopen{}\frac{\partial#1}{\partial#2}\mathclose{}}

\begin{document}

\title{Moir\'e-enabled chiral plasmons in twisted bilayer graphene nanoislands}

\author{Magnus~Husted\,\orcidlink{0009-0002-1483-6100}}
\affiliation{POLIMA---Center for Polariton-driven Light--Matter Interactions, University of Southern Denmark, Campusvej 55, DK-5230 Odense M, Denmark}

\author{Viktor~Kanstrup\,\orcidlink{0009-0007-0751-9111}}
\affiliation{POLIMA---Center for Polariton-driven Light--Matter Interactions, University of Southern Denmark, Campusvej 55, DK-5230 Odense M, Denmark}

\author{Eduardo~J.~C.~Dias\,\orcidlink{0000-0002-6347-5631}}
\email[Eduardo~J.~C.~Dias: ]{dias@mci.sdu.dk}
\affiliation{POLIMA---Center for Polariton-driven Light--Matter Interactions, University of Southern Denmark, Campusvej 55, DK-5230 Odense M, Denmark}

\author{Joel~D.~Cox\,\orcidlink{0000-0002-5954-6038}}
\email[Joel~D.~Cox: ]{cox@mci.sdu.dk}
\affiliation{POLIMA---Center for Polariton-driven Light--Matter Interactions, University of Southern Denmark, Campusvej 55, DK-5230 Odense M, Denmark}
\affiliation{Danish Institute for Advanced Study, University of Southern Denmark, Campusvej 55, DK-5230 Odense M, Denmark}

\begin{abstract}
Twisting two stacked graphene layers superimposes a moir\'e superlattice that reshapes their electronic structure, yet how it shapes the collective optical response of finite, nanoscale bilayers remains largely unexplored. Using an atomistic tight-binding and self-consistent density-matrix approach, we compute the twist-dependent plasmonic response of twisted bilayer graphene nanoislands comprising thousands of atoms. The spectra are strongly tunable and split into two regimes at a critical angle where a single moir\'e wavelength spans the island; beyond it, an ordered ladder of resonances emerges whenever an integer number of moir\'e periods fits within the overlapping bilayer. In doped islands, the interlayer coupling also produces a moir\'e-enabled Born--Kuhn chiral plasmon: opposite transverse dipoles in the two layers form an electric quadrupole that hybridizes with the bright dipole, opening an anticrossing near the critical angle, with a handedness that reverses with the twist direction. Triangular islands confirm that the moir\'e lattice, not the flake boundary, governs both effects, establishing twist as a mechanical handle on chiral graphene plasmons.
\end{abstract}

\maketitle

\section{Introduction}

Graphene---the atomically thin honeycomb allotrope of carbon---combines a gapless linear (Dirac) dispersion, a universal optical conductivity, and an electrically tunable response, making it a versatile platform for photonics and optoelectronics~\cite{novoselov2004electric,novoselov2005twodimensional,nair2008fine,bonaccorso2010graphene,cox2016quantum}. Stacking two graphene sheets introduces interlayer hybridization that reshapes the low-energy spectrum~\cite{lopesdossantos2007graphene,cao2018unconventional}, while imposing a relative twist $\theta$ between the layers superimposes a long-wavelength \textit{moir\'e} pattern of periodicity
\begin{equation} \label{eq:moire}
    \lambda_{\rm m}(\theta) = \frac{a}{2\sin(\theta/2)} ,
\end{equation}
where $a=0.246$\,nm is the graphene lattice constant~\cite{andrei2020graphene,koshino2020effective,dey2023atomistic}. Acting as a superlattice potential, the moir\'e modulation folds the bands into a mini-Brillouin zone and suppresses the Fermi velocity, yielding the flat bands that, near the so-called magic angle, host correlated insulating states and superconductivity~\cite{bistritzer2011moire,tramblydelaissardiere2012numerical,cao2018correlated,cao2018unconventional,kuang2025review}. Twisted bilayer graphene (TBG) has consequently emerged as a paradigmatic system in which the twist angle serves as a structural knob for engineering electronic structure on demand~\cite{andrei2020graphene,dey2023atomistic}.

Doped graphene also supports highly confined, long-lived, and gate-tunable plasmons across the infrared and terahertz that squeeze optical fields into deeply subwavelength volumes~\cite{garciadeabajo2014graphene}. Patterning graphene into nanoislands and related nanostructures discretizes these excitations into geometrically tunable localized plasmon resonances~\cite{thongrattanasiri2012quantum,cox2014electrically}. At few-nanometer scales, however, the response becomes governed by quantum finite-size effects---atomic-scale edge terminations, level quantization, and nonlocality---that classical electromagnetic treatments fail to capture and that instead demand an atomistic, quantum-mechanical description~\cite{thongrattanasiri2012quantum,silveiro2015quantum,cox2016quantum,garciadeabajo2025roadmap}.

These two themes---moir\'e band engineering and quantum-confined graphene plasmonics---have largely developed in isolation. Plasmons in \textit{extended} TBG have been predicted and imaged~\cite{stauber2016quasiflat,sunku2018photonic,hesp2021observation}, and moir\'e-related features have been identified in its absorption~\cite{moon2013optical}. For finite TBG nanoislands, however, most studies address single-particle absorption~\cite{tiutiunnyk2019opto,tepliakov2020twisted,wang2022polarizationdependent,magar2026ultrafast}, and discrete plasmonic excitations have so far been predicted only through EELS simulations of hexagonal TBG supercells for twist angles between $0\mydeg$ and $10\mydeg$~\cite{westerhout2022quantum}. Islands of a few nanometers across are smaller than the magic-angle moir\'e period ($\approx13$\,nm), so their response is set by the ratio between $\lambda_{\rm m}$ and the island size rather than by the twist angle alone. How this interplay between nanoscale confinement and moir\'e periodicity shapes the plasmonic response remains unknown.

\begin{figure*}[t]
    \centering
    \includegraphics[width=\linewidth]{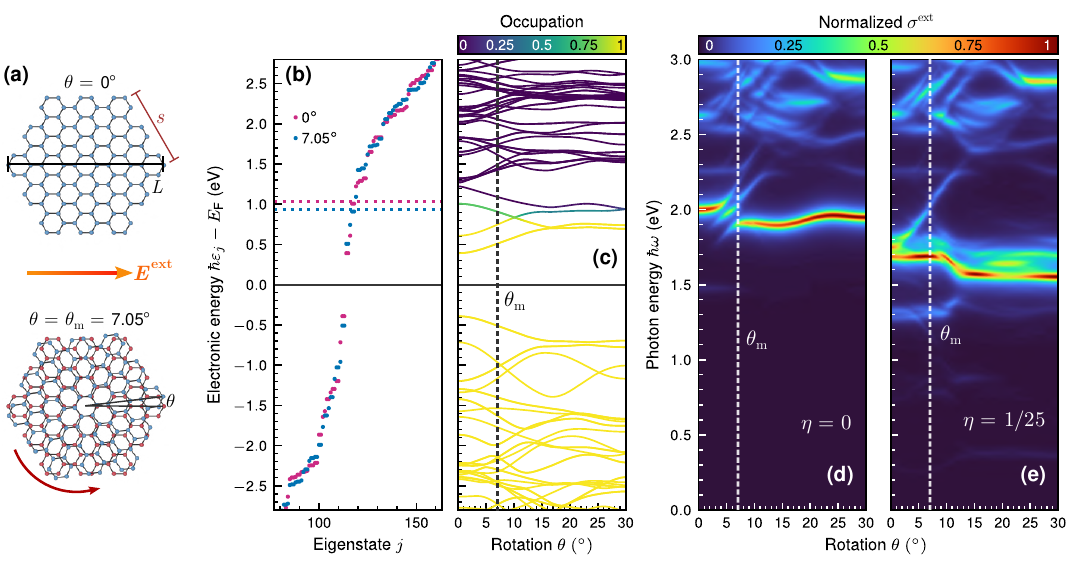}
    \caption{\textbf{Electronic and linear optical properties of twisted bilayer graphene nanoislands.} \textbf{(a)} Illustration of AA-stacked armchair edge-terminated hexagonal graphene nanoislands of tip-to-tip length $L=2$\,nm  (corresponding to 114 carbon atoms per layer) with relative twist angles $\theta=0\mydeg$ (upper schematic) and $\theta=\theta_{\text{m}}=$ $7.05\mydeg$ (lower schematic). \textbf{(b)} Eigenenergy states $\hbar\varepsilon_j$ (referenced to Fermi energy $\EF$) of the untwisted and twisted configurations ($\theta=\theta_{\rm m}$). The color-coded dotted lines (see legend) mark the doped chemical potential $\mu$ at the respective twist angles when the nanostructure is doped to $\eta=1/25$. \textbf{(c)} Eigenenergy states as a function of twist angle $\theta$, with symbol color depicting occupation level corresponding to a doping concentration of $\eta =1/25$, where the solid black line marks the Fermi energy of the undoped system (fully occupied states below, empty above). \textbf{(d,e)} Linear optical response of the twisted nanostructures, represented by the normalized extinction cross section, as a function of twist angle for the stacked hexagonal graphene nanoislands, for the (d) undoped and (e) doped ($\eta=1/25$) cases. The critical angle $\theta_{\text{m}}$ is marked by the vertical dashed lines.
    }
    \label{fig1}
\end{figure*}

A second, and so far largely separate, consequence of the twist is chirality. Because rotations by $\theta$ and $-\theta$ produce mirror-image stackings, TBG is intrinsically handed, and this handedness imprints an optical activity and chiral electronic response on the extended crystal~\cite{stauber2018chiral} that, at finite doping, can be inherited by its collective modes as chiral plasmons~\cite{huang2022observation}. In classical optics, the archetype of such behavior is the Born--Kuhn model: two vertically stacked, mutually rotated dipole oscillators whose coupling generates circular dichroism and optical rotation, and which underpins the modern design of chiral plasmonic metamolecules~\cite{yin2013interpreting}. A twisted bilayer nanoisland is a natural atomic-scale analogue of this geometry---two stacked, relatively rotated sheets of oscillating charge---in which handedness can arise both from the shape of the flakes and from the twisted atomic lattice itself. Whether the moir\'e lattice alone can produce a genuine Born--Kuhn chiral plasmon in a finite, quantum-confined TBG island, and how it governs such a mode, has not been established.

Here we bridge these gaps by computing, from an atomistic standpoint, the twist-dependent optical response of hexagonal and triangular TBG nanoislands, whose high symmetry suppresses shape-driven (dipolar) Born--Kuhn coupling and thus isolates the role of the lattice. Combining a tight-binding description of the electronic states with a self-consistent single-particle density-matrix formalism, we model structures comprising thousands of carbon atoms while retaining the edge and moir\'e physics inaccessible to continuum models~\cite{cox2014electrically,devega2020strong}. We report two intertwined findings. First, the twist angle strongly modulates both the electronic spectrum and the extinction cross section, and a critical angle---at which a single moir\'e wavelength first spans the island---separates two distinct spectral regimes; beyond it, an ordered ladder of resonances emerges whenever an integer number of moir\'e wavelengths fits within the overlapping bilayer region, linking the collective optical response directly to moir\'e commensurability. Second, and enabled by the same interlayer coupling, the doped twisted island develops a \emph{moir\'e-enabled Born--Kuhn chiral plasmon}: opposite transverse dipoles induced in the two layers form an electric quadrupole $Q_{yz}$ that hybridizes with the bright dipole, producing an anticrossing in the extinction spectra near the critical angle, while the cross product of the layer dipoles---a pseudoscalar that reverses sign under $\theta\to-\theta$---quantifies the handedness of the mode. The triangular geometry, whose three-fold flake symmetry ($D_{3h}$) is deliberately mismatched to the six-fold moir\'e pattern, confirms that both the resonance ladder and the chiral mode are dictated by the moir\'e lattice rather than the flake boundary, establishing twist as a mechanical route to chiral, geometrically tunable plasmons in graphene nanostructures.

\section{Results and discussion}

We model the optical response of TBG nanoislands atomistically, combining a tight-binding description of carbon $p_z$ electrons with a self-consistent single-particle density matrix formalism (see Methods) that captures edge, confinement, and interlayer moir\'e effects in structures of thousands of atoms. We focus on armchair edge terminations, which damp plasmons less strongly than zigzag ones~\cite{thongrattanasiri2012quantum,cox2014electrically}.

As the rotation angle $\theta$ between the two layers increases, the moir\'e wavelength shrinks according to Eq.~\eqref{eq:moire}, while the overlap region between the two layers also changes. We identify the critical (moir\'e commensurate) angle $\theta_{\rm m}$ as the angle at which one moir\'e wavelength spans the island. More generally, the commensurate angles $\theta_n$, at which $n$ moir\'e wavelengths first span the longest chord $D(\theta)$ of the overlap region (see Methods), follow from
\begin{align}
    n\lambda_{\text{m}}(\theta_{n}) = D(\theta_n) .
\end{align}
We use $D$ rather than the fixed tip-to-tip length $L$ because it sets the relevant length scale at large twist. At the small angles relevant for $\theta_{\rm m}$, $D$ differs from $L$ by only a few percent, and we define $\theta_{\rm m}$ through $\lambda_{\rm m}(\theta_{\rm m})=L$, so that $\theta_{\rm m}\approx\theta_1$. The commensurability condition is meaningful only where $\lambda_{\rm m}(\theta_n)\gg a$, so that enough atoms resolve the moir\'e pattern.

Figure~\ref{fig1} summarizes the effect of twist on the smallest hexagonal island ($L\approx 2$\,nm, $N=228$ atoms). A finite twist [Fig.~\ref{fig1}(a), shown for $\theta=\theta_{\rm m}=7.05\mydeg$] lowers the symmetry of the aligned bilayer (from $D_{6h}$ to $D_6$) and reorganizes the eigenstates [Fig.~\ref{fig1}(b)] through modified interlayer hybridization and confinement. Over the irreducible range $0$--$30\mydeg$ [Fig.~\ref{fig1}(c)], the spectrum evolves markedly: the undoped HOMO--LUMO gap widens from $0.78$ to $1.38$\,eV, and individual levels shift in energy and degeneracy, establishing the twist angle as an effective structural parameter for the electronic structure of the island.

The twist-induced electronic structure rearrangements translate into the carrier occupations upon doping: adding $N_{\rm dop}=\eta N$ carriers shifts the chemical potential to conserve charge [color scale in Fig.~\ref{fig1}(b,c)], and for $\eta=1/25$ at $T=300$\,K---corresponding to a $\sim1$\,eV shift accessible by electrostatic gating in extended graphene~\cite{chen2011controlling,fei2012gatetuning}---the Fermi level enters a high-density-of-states region where the twist-induced level reshuffling strongly modulates the occupations compared to the charge-neutral case.

\begin{figure*}[t]
    \centering
    \includegraphics[width=\linewidth]{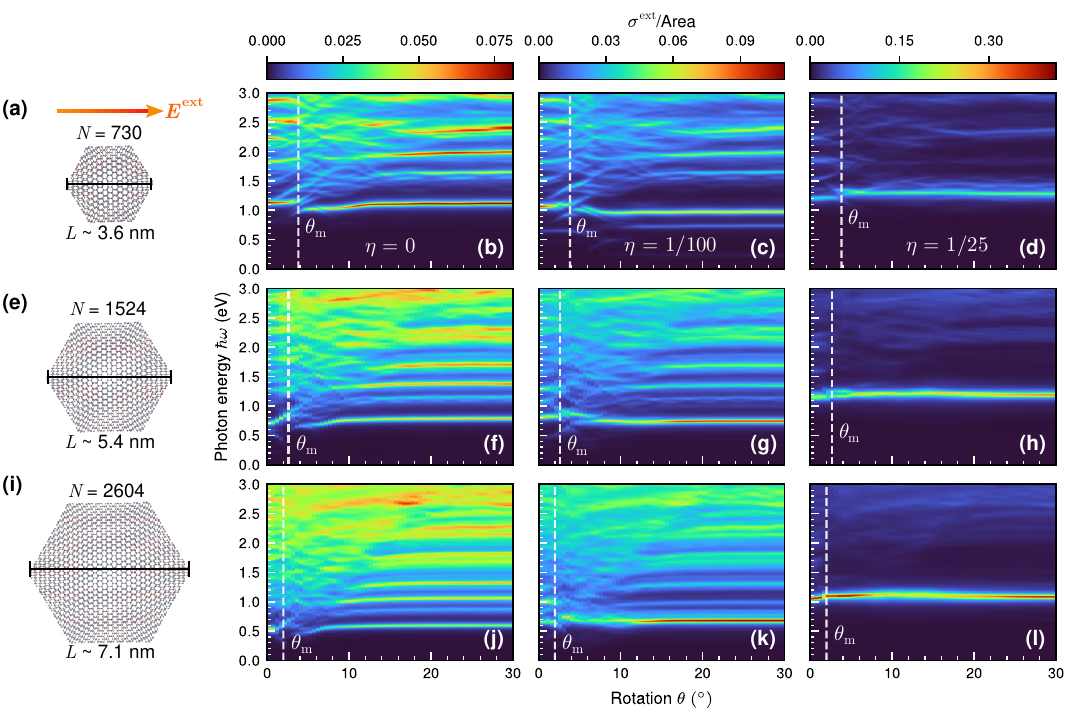}
    \caption{\textbf{Dependence of linear optical response of hexagonal TBG nanoislands on size and doping.} \textbf{(a,e,i)} atomic structures of the three different-sized TBG nanoislands: (a) $N=730$ ($L\approx3.6\,$nm), (e) $N=1524$ ($L\approx5.4\,$nm), and (i) $N=2604$ ($L\approx7.1\,$nm). The polarization direction of the external field $E^{\text{ext}}$ is indicated by the arrow. \textbf{(b--d,f--h,j--l)} Extinction cross section $\sigma^{\text{ext}}$, normalized to the island area, as a function of twist angle $\theta$ and photon energy $\hbar\omega$, for (b--d) the small, (f--h) medium, and (j--l) large systems, and for doping levels of (b,f,j) $\eta=0$ (undoped case), (c,g,k) $\eta=1/100$, and (d,h,l) $\eta=1/25$. The dashed line in each panel marks the critical angle $\theta_{\text{m}}$ for the corresponding nanoisland size.
    }
    \label{fig2}
\end{figure*}

This electronic tunability carries over to the optical response. The extinction cross section $\sigma^{\rm ext}$ under an $x$-polarized field [Fig.~\ref{fig1}(d,e); see also Methods] features a single dominant resonance near $2.0$\,eV in the undoped island and near $1.7$\,eV at $\eta=1/25$, where the response acquires an increasingly collective, plasmonic character; in both cases the resonance is strongly modulated by twist, more so under doping. Most strikingly, the spectra split into two regimes about $\theta_{\rm m}$. In the undoped island, the dominant resonance fades at $\theta_{\rm m}$ and a red-shifted peak emerges; because this discontinuity vanishes when electron--electron interactions are switched off (see Fig.~\ref{figS1} in the Appendix), it reflects a collective response to the moir\'e-driven reconfiguration of the electronic states. The doped island likewise red-shifts beyond $\theta_{\rm m}$ and remains strongly angle dependent.

\begin{figure*}[t!]
    \centering
    \includegraphics[width=\linewidth]{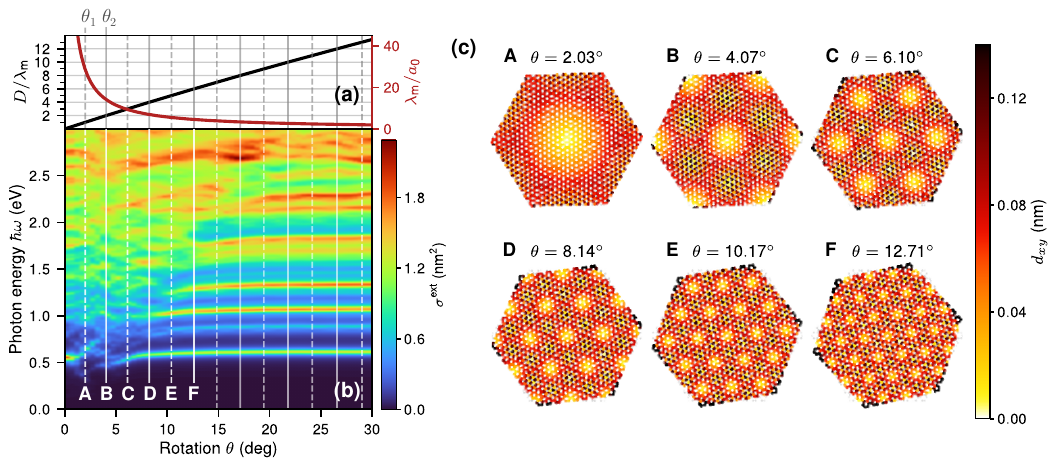}
    \caption{\textbf{Effect of moir\'e commensurability on the linear optical response of hexagonal TBG nanoislands.} \textbf{(a)} Number of moir\'e wavelengths $\lambda_{\text{m}}$ supported along the longest overlap chord $D$ (black, left axis) and moir\'e wavelength (red, right axis) for two hexagonal graphene nanoislands ($N=2604$, $L=7.1\,$nm), as a function of twist angle $\theta$. Intersections of this curve with integer values $n$ are marked by dashed (solid) vertical lines identifying odd (even) values of $n$. \textbf{(b)} Extinction cross section of the undoped island as a function of photon energy $\hbar\omega$ and twist angle $\theta$. The vertical lines are the same as in (a), marking integer values of supported moir\'e wavelengths. \textbf{(c)} TBG nanoisland atomic positions at the marked $\theta_{n} $ positions on (b), labeled A--F, with the in-plane projection distance $d_{xy}$ between the top layer atoms and bottom layer atoms indicated by the colormap.}
    \label{fig3}
\end{figure*}

The behavior in Fig.~\ref{fig1} suggests that the critical angle $\theta_{\rm m}$ organizes the twist-dependent optical response more generally. To test this across length scales and carrier doping densities, we examine three hexagonal armchair nanoislands---$N=730$ ($L\approx 3.6$\,nm), $N=1524$ ($L\approx 5.4$\,nm), and $N=2604$ ($L\approx 7.1$\,nm), hereafter small, medium, and large---at dopings $\eta=0$, $1/100$, and $1/25$ [Fig.~\ref{fig2}]. The extinction cross section, normalized to the island area, is mapped over the irreducible range $0$--$30\mydeg$ and an energy window comparable to the hopping amplitude $t_0=2.8$\,eV. A change of regime at $\theta_{\rm m}$ (dashed lines) persists across every size and doping. This phenomenon is clearest in the undoped maps [Figs.~\ref{fig2}(b,f,j)], where several near-flat modes for $\theta<\theta_{\rm m}$ reorganize sharply above $\theta_{\rm m}$, settling into an ordered ladder of nearly flat resonances, with higher-energy rungs setting in at progressively larger angles. The ladder sharpens with island size, as larger structures support more modes.

Doping preserves this picture but progressively obscures it: the lowest-order mode brightens steeply with $\eta$ (see Fig.~\ref{figS2} in the Appendix) and dominates the normalized maps, while its energy increases with doping (from $\approx0.6$\,eV at $\eta=0$ to $\approx1.1$\,eV at $\eta=1/25$ in the large island)---the signature of an emerging plasmon~\cite{devega2020strong}. This plasmon shifts abruptly near $\theta_{\rm m}$ and stabilizes at larger angles. Incidentally, because AB-stacked islands behave qualitatively the same (see Fig.~\ref{figS3} in the Appendix), we present AA stacking throughout.

While the critical angle accounts for the abrupt reorganization at $\theta_{\rm m}$, it does not explain the flat spectral features of the undoped islands in Fig.~\ref{fig2}. To test whether these are tied to the moir\'e geometry, we focus on the largest hexagonal island ($N=2604$), which supports the most resonances. Figure~\ref{fig3}(a) plots the number of moir\'e wavelengths spanning the longest overlap chord, $D/\lambda_{\rm m}$, versus twist angle [$D$ from Eq.~\eqref{eq:D_hex}]; each crossing of an integer $n$ (solid lines for even $n$, dashed for odd) defines a commensurate angle $\theta_n$ beyond which the overlap region accommodates at least $n$ moir\'e wavelengths. Overlaying these lines on the extinction map of Fig.~\ref{fig2}(j), reproduced in Fig.~\ref{fig3}(b), links the optical response to the finite-size moir\'e pattern, which we visualize at the marked angles A--F in Fig.~\ref{fig3}(c) by coloring top-layer atoms according to their in-plane offset $d_{xy}$ from the bottom layer (bright: AA alignment; dark: AB/BA).

In Fig.~\ref{fig3}(b), the boundary between the dispersive low-angle response and the ordered ladder of flat resonances runs diagonally: higher-energy rungs set in at larger twist, in step with the growth of $D/\lambda_{\rm m}$ in Fig.~\ref{fig3}(a). The ladder thus emerges as successive integer numbers of moir\'e wavelengths become commensurate with the island. The atomic maps confirm this progression: at $n=1$ (A) a single central AA spot forms, surrounded by AA regions emerging at the edges; at $n=2$ (B) this spot shrinks as six corner spots appear, fully resolving at $n=3$ (C); and for $n=4$--$6$ (D--F) progressively more, smaller AA spots arrange into the characteristic hexagonal moir\'e superlattice.

\begin{figure*}[t!]
    \centering
    \includegraphics[width=\linewidth]{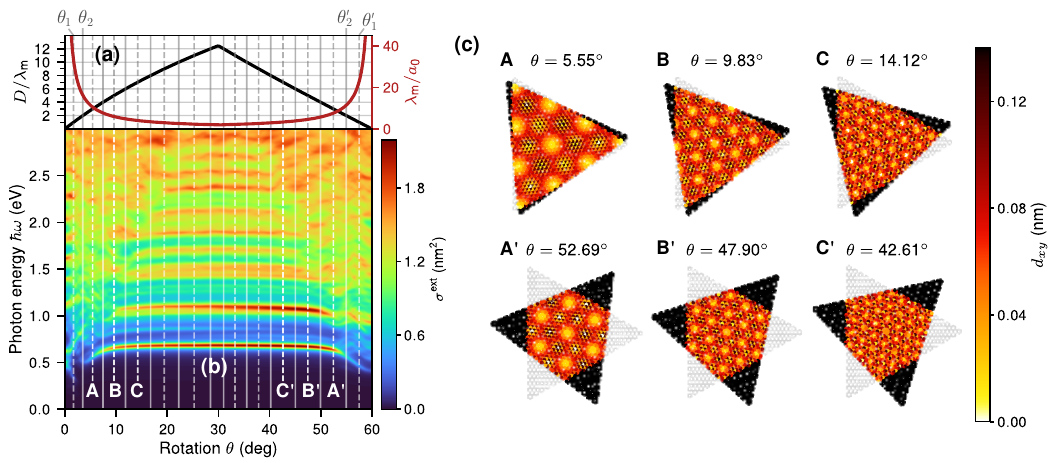}
    \caption{\textbf{Effect of moir\'e commensurability on the linear optical response of triangular TBG nanoislands.} \textbf{(a)} Number of moir\'e wavelengths $\lambda_{\rm m}$ supported along the longest overlap chord $D$ (black, left axis) and moir\'e wavelength (red, right axis) for two triangular graphene nanoislands ($N=2520$, height $L=7.26$\,nm) as a function of twist angle $\theta$. Intersections of this curve with integer values $n$ are marked by dashed and solid lines, representing odd and even values of $n$ respectively, where the hexagonal symmetry of $\lambda_{\rm m}$ folds the curve about $\theta=30\mydeg$, so that $\lambda_{\rm m}(\theta_n)=\lambda_{\rm m}(60\mydeg -\theta_n)$. \textbf{(b)} Extinction cross section of the undoped triangular TBG nanoislands as a function of energy $\hbar\omega$ and twist angle $\theta$. The vertical lines are the same as in (a). \textbf{(c)} TBG nanoisland atomic positions at the marked $\theta_n$ positions on (b) labeled A--C and A'--C' for $\lambda_{\rm m}$-equivalent angles, with the in-plane projection distance $d_{xy}$ between the top layer atoms and bottom layer atoms indicated by the colormap.}
    \label{fig4}
\end{figure*}

The flat modes track this sequence: the lowest (just above $0.5$\,eV) sets in at $\theta_2$ (B) and intensifies at $\theta_3$ (C), a second ($\approx0.8\,$eV) and third ($\approx 1.1$\,eV) appear together at $\theta_4$ (D), and a fourth ($\approx 1.3$\,eV) emerges at $\theta_5$ (E) and sharpens at $\theta_6$ (F), while higher orders grow increasingly diffuse. This correspondence is qualitative, offering one interpretation of the link between moir\'e commensurability and the observed flat ladder modes. Commensurability-dependent absorption has been reported for \textit{extended} TBG~\cite{moon2013optical}, but the interplay between overlap size and the confined moir\'e periodicity seen here appears distinct, motivating a similar analysis for the triangular geometries presented in Fig.~\ref{fig4}.

We corroborate this picture with triangular nanoislands of comparable size ($N=2520$), whose three-fold flake symmetry ($D_{3h}$) is deliberately mismatched to the six-fold symmetry of the moir\'e pattern. Figure~\ref{fig4}(a) shows $D/\lambda_{\rm m}$ across the full $0$--$60\mydeg$ irreducible range [$D$ from Eq.~\eqref{eq:D_tri}]. Because the moir\'e wavelength inherits the six-fold symmetry of the lattice, it folds about $\theta=30\mydeg$, $\lambda_{\rm m}(\theta)=\lambda_{\rm m}(60\mydeg-\theta)$~\cite{lopesdossantos2007graphene}, whereas the overlap chord $D$ does not: a $60\mydeg$ rotation inverts the triangle, so the geometries at $\theta$ and $60\mydeg-\theta$ differ. The commensurability condition therefore recurs on the descending branch at angles $\theta_n'$ that share the order $n$ of their ascending-branch partners $\theta_n$ but realize it with a different overlap geometry. The atomic maps [Fig.~\ref{fig4}(c)] bear this out: from $\theta_3$ (A) to $\theta_5$ (B) to $\theta_7$ (C) the AA spots shrink to fit more wavelengths, while past $30\mydeg$ the same orders reappear at C'--A' with the spots growing again.

The flat ladder modes of the hexagonal islands [Fig.~\ref{fig3}(b)] reappear in the triangular extinction map [Fig.~\ref{fig4}(b)], now evolving symmetrically about $\theta=30\mydeg$---each rung emerges on the ascending branch and fades in reverse order on the descending one. This symmetry is the key advantage of the triangular geometry: because the modes follow the six-fold symmetry of the moir\'e pattern rather than the three-fold symmetry of the flake, the commensurate moir\'e orders, and not the flake boundary, emerge as the organizing principle behind the ladder. The precise energies and intensities of the modes are further shaped by island size and edge termination, as shown for zigzag edges in Fig.~\ref{figS4} of the Appendix.

\begin{figure*}[t!]
    \centering
    \includegraphics[width=\linewidth]{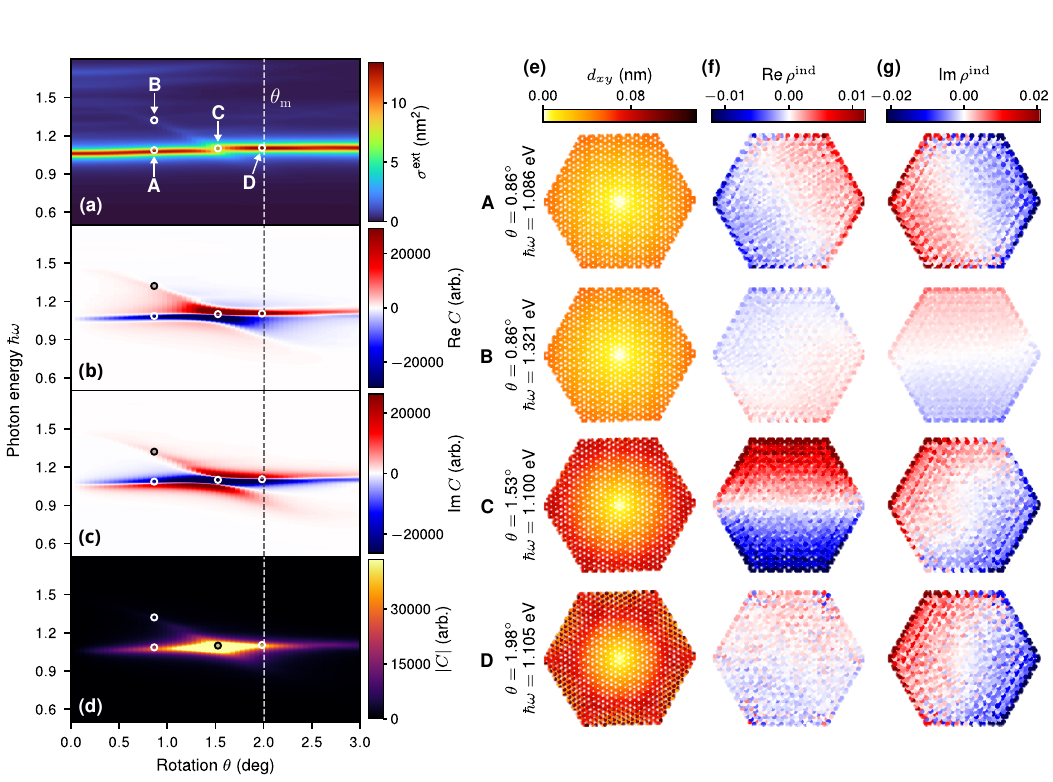}
    \caption{\textbf{Moir\'e-enabled Born--Kuhn chiral plasmon in hexagonal TBG nanoislands.} \textbf{(a)} Extinction cross section of the highly doped ($\eta=1/25$) large hexagonal TBG nanoislands ($L=7.1$\,nm, $N=2604$) as a function of energy $\hbar\omega$ and twist angle $\theta$. Points A--D mark corresponding angles and energies at which the frequency-domain induced charge density $\rho^{\mathrm{ind}}$ is presented. The dashed line marks the critical angle $\theta_{\mathrm{m}}$. \textbf{(b)}--\textbf{(d)} The real part, imaginary part and modulus of the Born--Kuhn chirality parameter $C$ respectively (in arbitrary units). \textbf{(e)} TBG nanoisland atomic positions at corresponding A--D positions in (a), with the in-plane projection distance $d_{xy}$ between top and bottom atoms indicated via the colormap. \textbf{(f)} Real part of the frequency-domain induced charge density per atom of the top layer at A--D marked positions in (a). \textbf{(g)} Same as in (f) but for the imaginary part.}
    \label{fig5}
\end{figure*}

\textit{Moir\'e-enabled Born--Kuhn chiral plasmon.} The commensurability analysis above concerns the undoped islands, where the ladder reflects the moir\'e-driven reconfiguration of electronic states. In the doped, collective regime the same interlayer coupling has a second consequence: it couples the bright plasmon to a mode that is dark in the aligned bilayer. Figure~\ref{fig5}(a) maps the extinction of the highly doped ($\eta=1/25$) large hexagonal island at small twist angles. Rather than shifting monotonically, the dominant plasmon near $1.1$\,eV is met, just below $\theta_{\rm m}$ (dashed line), by a second branch that descends from higher energy as the twist increases, and the two repel in an anticrossing [points A--D], across which the bright plasmon shifts upward in energy. The second branch is faint in the extinction but stands out clearly in the chirality maps of Fig.~\ref{fig5}(b--d), as expected for a quadrupolar mode. This is dipole--quadrupole hybridization: the bright dipole, in which the two layers oscillate in phase, couples to a layer-antisymmetric partner. In the aligned bilayer the horizontal mirror plane forbids their mixing; the twist removes it ($D_{6h}\to D_6$), allowing the two to hybridize.

That the hybridized plasmon is chiral follows from the Born--Kuhn parameter $C(\ww)$ [see Eq.~\eqref{eq:C} in Methods], shown in Fig.~\ref{fig5}(b--d). By symmetry, $C$ vanishes for the aligned bilayer and reverses sign under $\theta\to-\theta$, so a finite value directly signals the handedness of the induced charge distribution. Its modulus $|C|$ [Fig.~\ref{fig5}(d)] grows with twist, peaks at the anticrossing (point C), and is suppressed away from it: the layer dipoles become markedly non-collinear precisely where the bright dipole hybridizes with its layer-antisymmetric partner, as expected for coupled Born-Kuhn oscillators. The sign changes of $\Ree\{C\}$ and $\Imm\{C\}$ across the resonance [Fig.~\ref{fig5}(b,c)] reflect the relative phase of $p_x$ and $Q_{yz}$ (see Methods) rather than a change of handedness.

Microscopically, the twist-induced chirality is written into the induced charge. Near the anticrossing, the top-layer induced charge density [Fig.~\ref{fig5}(f,g)] develops a dipolar profile with a marked transverse ($y$) component that is absent at $\theta=0$; at point C it is almost entirely transverse, as is the upper branch at point B. Displaced vertically from its oppositely directed counterpart in the bottom layer, this transverse layer dipole is exactly the $Q_{yz}$ quadrupole of Eq.~\eqref{eq:Qyz}, and together with the bright dipole $p_x$ it yields the finite $C$ of Fig.~\ref{fig5}(b--d) [Eq.~\eqref{eq:C_Q}]. The chiral plasmon is therefore not imposed by the flake boundary but generated internally, by the twist-dependent interlayer coupling [Fig.~\ref{fig5}(e)]. Notably, the anticrossing occurs for $\theta\lesssim\theta_{\rm m}$, where the island samples less than one moir\'e period: the chirality is set by the variation of the atomic registry across the island rather than by a fully developed superlattice.

\begin{figure*}[t!]
    \centering
    \includegraphics[width=\linewidth]{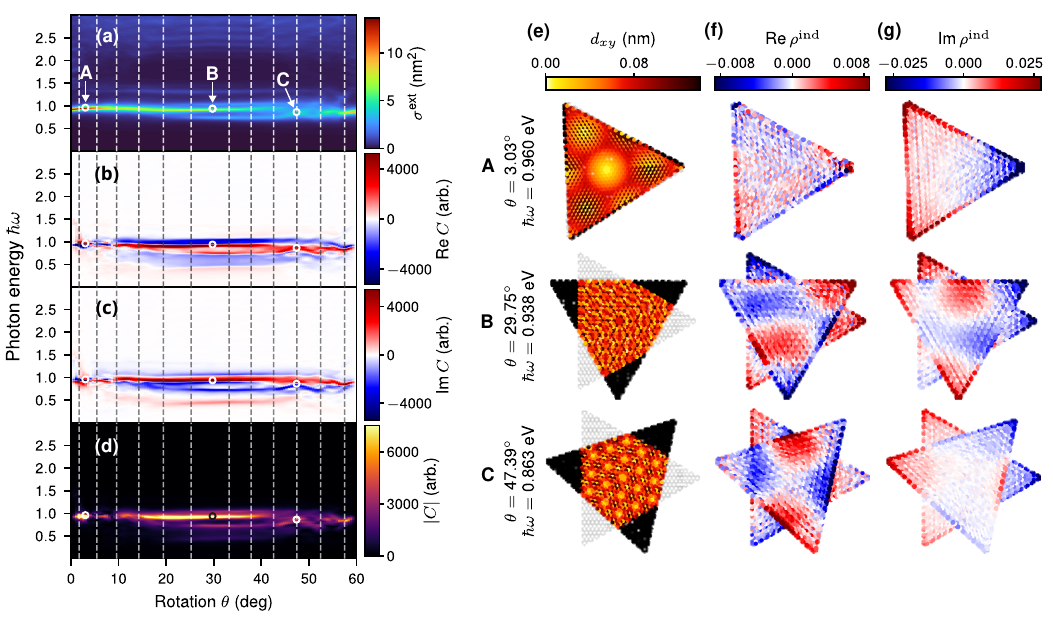}
    \caption{\textbf{Lattice-symmetry control of the Born--Kuhn chiral plasmon in triangular TBG nanoislands.} \textbf{(a)} Extinction cross section of the highly doped ($\eta=1/25$) large triangular TBG nanoislands ($L=7.26$\,nm, $N=2520$) as a function of energy $\hbar\omega$ and twist angle $\theta$. Points A--C mark corresponding angles and energies at which the frequency-domain induced charge density $\rho^{\mathrm{ind}}$ is presented. Dashed lines mark odd values of the commensurate order $n$ supported across the overlap region. \textbf{(b)}--\textbf{(d)} The real part, imaginary part and modulus of the Born--Kuhn chirality parameter $C$ respectively. \textbf{(e)} TBG nanoisland atomic positions at corresponding A--C positions in (a), with the in-plane projection distance $d_{xy}$ between top and bottom atoms indicated via the colormap. \textbf{(f)} Real part of the frequency-domain induced charge density per atom at A--C marked positions in (a). \textbf{(g)} Same as in (f) but for the imaginary part.}
    \label{fig6}
\end{figure*}

The triangular islands confirm that the chiral mode, like the resonance ladder, is organized by the moir\'e lattice rather than the flake, but with a markedly softer angular dependence. In the doped triangular island [$\eta=1/25$, Fig.~\ref{fig6}], the chirality parameter $C$ is not concentrated near a single anticrossing as in the hexagons but is distributed across a broad range of twist angles. We attribute this to the interlayer overlap of two co-centered triangles being both smaller and far more strongly angle-dependent than that of two hexagons: whereas rotated hexagons retain more than $90\%$ of their overlap area across the full irreducible range, the triangular overlap falls monotonically to two-thirds at $\theta=60\mydeg$. Fewer atoms therefore lie in the bilayer region that sustains the interlayer coupling, so the commensurability condition is less sharply resolved and the chiral response spreads over angle. It nonetheless remains lattice-governed---its envelope is approximately symmetric about $\theta=30\mydeg$, following the six-fold symmetry of the moir\'e wavelength rather than the three-fold symmetry of the flake (points A--C mark representative angles). Together with the reversal $C(-\theta)=-C(\theta)$, this establishes a moir\'e-enabled chiral plasmon whose handedness is set by twist, rendering the twist angle a mechanical control over chiral light--matter interaction in an atomically thin, all-carbon platform.

\section{Conclusions}

In summary, we have used an atomistic tight-binding and self-consistent density-matrix framework---capturing edge, confinement, and interlayer moir\'e physics in TBG nanoislands comprising thousands of atoms---to show that the interlayer twist governs both the collective optical spectrum and its chirality. The twist angle strongly modulates the electronic structure and extinction, with a critical angle, at which one moir\'e wavelength first spans the island, separating two spectral regimes; beyond it, an ordered ladder of flat resonances tracks the successive commensurability of integer numbers of moir\'e periods with the overlapping bilayer. In doped islands, the same interlayer coupling induces opposite transverse dipoles in the two layers, forming an electric quadrupole $Q_{yz}$ that hybridizes with the bright dipole. The result is a moir\'e-enabled Born--Kuhn chiral plasmon that anticrosses just below the critical angle, where the island samples less than one moir\'e period, and with handedness that reverses under $\theta\to-\theta$. Triangular islands, the three-fold flake symmetry of which is deliberately mismatched to the six-fold moir\'e pattern, confirm that the moir\'e lattice rather than the flake boundary organizes both the resonance ladder and the chiral mode, the latter spread over a broader angular range by the smaller interlayer overlap of the triangular geometry.

These results establish the interlayer twist as a purely mechanical route to chiral, geometrically tunable plasmons in an atomically thin, all-carbon platform, complementing the chiral responses reported for \textit{extended} TBG~\cite{kim2016chiral,stauber2018chiral,huang2022observation} by extending them to the quantum-confined, finite-size regime. Several directions follow naturally. Larger islands would resolve additional commensurate orders at progressively smaller angles and sharpen the chiral anticrossings; the doping dependence of the response points to an electrically tunable chiral plasmon; and computing the circular dichroism from the layer-resolved response (see Methods) would convert the chirality diagnosed here into a direct far-field observable. The strong, characteristic twist dependence also suggests twist-angle spectroscopic fingerprinting~\cite{moon2013optical}. Finally, lattice relaxation, which we have neglected and which is expected to be weak at these sizes~\cite{nam2017lattice,wang2022polarizationdependent}, should be examined at the small twist angles where the chiral anticrossing occurs.

\section{Methods}

Our approach couples an atomistic tight-binding description of the electronic states in TBG nanoislands to a self-consistent single-particle density-matrix treatment of their optical response, enabling simulations of structures with thousands of atoms while retaining edge, confinement, and interlayer moir\'e physics. We describe, in turn, the nanoisland geometry, the moir\'e commensurability analysis, the tight-binding Hamiltonian, and the density-matrix formalism from which the extinction and the layer-resolved multipole moments are obtained.

\subsection{Geometry}

\begin{figure}[bt]
    \centering
    \includegraphics[width=1.0\linewidth]{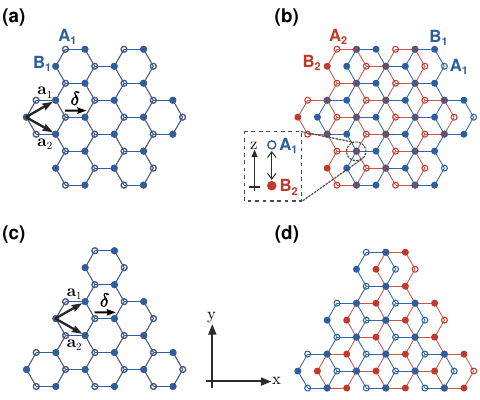}
    \caption{\textbf{Bilayer graphene nanostructure geometries.} \textbf{(a)} AA-stacked hexagonal nanoisland with armchair edge terminations; filled and unfilled circles denote A and B sublattice sites, respectively. Unit and displacement vectors indicated by arrows. \textbf{(b)} AB-stacked hexagonal nanoisland with armchair edges; upper and lower layers are shown in blue and red. Inset: the B sublattice of the upper layer sits directly above the A sublattice of the lower layer, characteristic of Bernal stacking. \textbf{(c,d)} Triangular nanoislands (rotated $30\mydeg$) with armchair edges in (c) AA and (d) AB stacking, following the same color conventions as (a) and (b).}
    \label{fig7}
\end{figure}

We build TBG nanoislands by stacking two graphene layers, each with sublattices A$_i$, B$_i$ ($i=1,2$) set by lattice vectors $\mathbf{a}_1=a_0 (3/2, \sqrt{3}/2)$, $\mathbf{a}_2=a_0(3/2, -\sqrt{3}/2)$ and basis offset $\delta = (-a_0,0)$, with $a_0=0.142$\,nm~\cite{castroneto2009electronic, mirzakhani2025magnetism, yoshizawa2000orbital}. Replicating the unit cell yields hexagonal or triangular flakes with armchair or zigzag edges (Fig.~\ref{fig7}), stacked a distance $b_0=0.335$\,nm apart (the graphite interlayer spacing)~\cite{wang2022polarizationdependent, tramblydelaissardiere2012numerical} to give $N$ atoms at positions $\rb_l$. We treat both AA and AB (Bernal) stacking, distinguished by whether A$_1$ lies above A$_2$ or B$_2$ (Fig.~\ref{fig7}). A twist is applied by rotating one layer about the common centroid over the irreducible range ($0$--$30\mydeg$ for hexagons, $0$--$60\mydeg$ for triangles; AB stacking gives the same ranges). The twist removes all mirror planes, reducing the symmetry of AA-stacked hexagonal (triangular) islands from $D_{6h}$ ($D_{3h}$) to $D_6$ ($D_3$), so that islands twisted by $\theta$ and $-\theta$ are mirror images. We assume rigid rotations: relaxation only weakly affects the optical response at these sizes~\cite{wang2022polarizationdependent}, and domain reconstruction is significant only below $2\mydeg$~\cite{nam2017lattice}.

\subsection{Moir\'e commensurability}

The number of moir\'e wavelengths spanning the longest overlap chord is $n(\theta)=D(\theta)/\lambda_{\rm m}(\theta)$, with $\lambda_{\rm m}$ from Eq.~\eqref{eq:moire} [with $\theta\to60\mydeg-\theta$ for $\theta>30\mydeg$] and integer $n$ defining the commensurate angles $\theta_n$.
For two overlapping triangles (side $s$) and hexagons (edge $s$),
\begin{equation}
    D_{\text{tri}} =
    \begin{cases}
        \dfrac{s}{2\sin{(30\mydeg+\theta/2)}},
            & 0 \leq \theta \leq 30\mydeg, \\[1.5ex]
        \dfrac{s}{2}\dfrac{\sin{(3\theta/2)}}{\sin{\theta}\sin{(\theta+60\mydeg)}},
            & 30\mydeg \leq \theta \leq 60\mydeg,
    \end{cases}
    \label{eq:D_tri}
\end{equation}
and
\begin{equation}
    D_{\text{hex}} = \frac{\sqrt{3}s}{\cos{(30\mydeg -\theta/2)}} ,
    \label{eq:D_hex}
\end{equation}
respectively, obtained as the largest vertex-to-vertex distance across the polygon formed by the overlap of the two rotated flakes.

\subsection{Electronic states}

Carbon $p_z$-orbitals $\ket{l}$ at sites $\rb_l$ are described by the tight-binding Hamiltonian
\begin{equation} \label{eq:H_TB}
    \Hm^{\rm TB} = \sum_{ll'}t_{ll'}\ket{l}\bra{l'} ,
\end{equation}
with hopping amplitudes of Slater--Koster form~\cite{tramblydelaissardiere2012numerical,wang2022polarizationdependent,kuang2025review},
\begin{equation}
    t_{ll'} = \beta^2V_{pp\sigma}(\rb_{ll'}) + \left(1-\beta ^2\right)V_{pp\pi}(\rb_{ll'}) ,
\end{equation}
where $\beta=\rb_{ll'}\cdot\zz/r_{ll'}$ is the direction cosine of $\rb_{ll'}\equiv\rb_{l'}-\rb_l$ along $z$, and
\begin{subequations}
\begin{align}
    V_{pp\pi}(\rb_{ll'}) &= -t_0\ee^{q\left(a_0-r_{ll'}\right)} F\left(r_{ll'}\right) ,
    \\
    V_{pp\sigma}(\rb_{ll'}) &= t_1\ee^{q\left(b_0-r_{ll'}\right)} F\left(r_{ll'}\right) ,
\end{align}
\end{subequations}
with $t_0=2.8$\,eV, $t_1=0.48$\,eV, $q=22.18\,{\rm nm}^{-1}$, and a smooth intra/inter-layer cutoff $F(r)=[1+\ee^{(r-r_{\rm c})/l_{\rm c}}]^{-1}$ ($r_{\rm c}=0.5$\,nm, $l_{\rm c}=0.0265$\,nm)~\cite{tramblydelaissardiere2012numerical,wang2022polarizationdependent}.

Diagonalizing Eq.~\eqref{eq:H_TB} yields the energies $\hbar\vep_j$ and coefficients $a_{jl}$ characterizing the electronic states $\ket{j}=\sum_l a_{jl}\ket{l}$, populated by the Fermi--Dirac distribution $f_j = \left[\ee^{(\hbar\vep_j-\mu)/\kB T}+1\right]^{-1}$. For a finite temperature $T$, the chemical potential $\mu$ follows from charge conservation, $N_{\rm e}+N_{\rm dop}=2\sum_j f_j$, where $N_{\rm e}=N$ is the number of electrons at charge neutrality and $N_{\rm dop}=\eta N$ sets the doping and the factor of two accounts for spin; at charge neutrality $\mu$ reduces to the Fermi energy, placed at the HOMO--LUMO midpoint so that it varies smoothly with temperature.

\subsection{Optical response}
\label{subsect:optical}

The optical response of a TBG nanoisland is simulated by considering its interaction with an external electric field $\Eb(t)$, quantified in the quasistatic approximation by the scalar potential $\phiext_l = -\rb_l\cdot\Eb$ at the carbon site $l$. To account for incoherent scattering effects, we characterize the state of a TBG nanoisland by the density matrix $\rho=\sum_{jj'}\rho_{jj'}\ket{j}\bra{j'}$, where $\rho_{jj'}$ are time-dependent matrix elements. At equilibrium, the density matrix is defined by $\rho_{jj'}^0=f_j\delta_{jj'}$. The induced charge $\rhoind_l$ is obtained from the diagonal elements of the density matrix in the site basis $\ket{l}$ according to $\rhoind_l=-2e(\rho_{ll}-\rho_{ll}^0)$, with $e$ denoting the elementary charge.

The induced charge sets up a potential $\phiind_l=\sum_{l'}v_{ll'}\rhoind_{l'}$ associated with electron--electron interactions within the nanoisland, with $v_{ll'}$ quantifying the spatial dependence of Coulomb coupling between orbitals at $\rb_l$ and $\rb_{l'}$. In practice, the elements of $v_{ll'}$ associated with in-plane electron--electron interactions in a TBG island are computed using a numerically fitted function for $p_z$ orbitals in graphene~\cite{thongrattanasiri2012quantum}, whereas the inter-plane interactions are modeled through a simple $1/r$ dependence.

The time evolution of the density matrix is governed by the Liouville--von Neumann equation~\cite{cox2014electrically,cox2017plasmonassisted,cox2019singleplasmon,devega2020strong}
\begin{equation}
    \pd{\rho}{t} = -\frac{\ii}{\hbar}\left[\Hm^{\rm TB}-e(\phiext+\phiind),\rho \right] - \frac{\gamma}{2}\left(\rho -\rho^0\right) ,
\end{equation}
with the first term capturing the tight-binding and electrostatic interactions and the second term accounting for relaxation to the equilibrium state through inelastic scattering, described by a phenomenological damping $\hbar\gamma=50$\,meV. Self-consistency enters through the induced potential~\cite{cox2014electrically,cox2017plasmonassisted,cox2019singleplasmon,devega2020strong}; we work in the linear-response regime, using a delta-like pulse polarized along $\xx$ to obtain the polarizability $\alpha(\ww)=\int {\rm d}t p_x(t)\ee^{\ii\ww t}$ from the induced dipole $\pb(t)=\sum_l\rhoind_l\rb_l$, from which the extinction cross section follows as $\sigma^{\rm ext}=(4\pi\ww/c)\Imm\{\alpha(\ww)\}$ (equal to absorption in this size regime)~\cite{devega2020strong,jelver2024nonlinear}.

\subsection{Layer-resolved dipoles and Born--Kuhn chirality}

Fourier transforming the induced charge, $\rhoind_l(\ww)=\int{\rm d}t\rhoind_l(t)\ee^{\ii\ww t}$, we resolve the induced dipole into contributions from the top (${\rm t}$) and bottom (${\rm b}$) sheets, located at $z_l=\pm b_0/2$ about the bilayer midplane,
\begin{equation}
    \pb_{\rm t}(\ww)=\sum_{l\in{\rm t}}\rhoind_l(\ww)\rb_l , \qquad \pb_{\rm b}(\ww)=\sum_{l\in{\rm b}}\rhoind_l(\ww)\rb_l .
\end{equation}
Each layer of a hexagonal or triangular island is three-fold symmetric, so its in-plane polarizability is isotropic: unlike the anisotropic oscillators of the classical Born--Kuhn model~\cite{yin2013interpreting}, a rigid rotation of an isolated layer leaves its response unchanged.
Chirality instead enters through the interlayer coupling---hopping and Coulomb interaction---which is sensitive to the relative atomic registry of the twisted lattices and can induce transverse, layer-antisymmetric dipoles. We quantify the resulting non-collinearity of the layer dipoles through the Born--Kuhn pseudoscalar
\begin{equation} \label{eq:C}
    C(\ww)=\zz\cdot\left[\pb_{\rm t}(\ww)\times\pb_{\rm b}(\ww)\right] = p_{{\rm t},x}p_{{\rm b},y}-p_{{\rm t},y}p_{{\rm b},x} .
\end{equation}

The vertical layer separation links $C$ directly to the electric quadrupole of the island, $Q_{ij}(\ww)=\sum_l \rhoind_l(\ww)\,r_{l,i}r_{l,j}$. Because $z_l=\pm b_0/2$,
\begin{equation} \label{eq:Qyz}
    Q_{yz}=\frac{b_0}{2}\left(p_{{\rm t},y}-p_{{\rm b},y}\right), \qquad Q_{xz}=\frac{b_0}{2}\left(p_{{\rm t},x}-p_{{\rm b},x}\right) ,
\end{equation}
so that, with the total in-plane dipole $\pb=\pb_{\rm t}+\pb_{\rm b}$,
\begin{equation} \label{eq:C_Q}
    C=-\frac{1}{b_0}\left(p_x Q_{yz}-p_y Q_{xz}\right) \xrightarrow[\;x\text{-pol.}\;]{}\; -\frac{1}{b_0}\,p_x Q_{yz} .
\end{equation}
The Born--Kuhn cross product is thus the product of the bright dipole $p_x$ and the quadrupole $Q_{yz}$, the interlayer-antisymmetric combination of the layer dipoles. By the three-fold symmetry of the islands, $Q_{yz}^{(x)}$ is the antisymmetric part of the tensor relating $(Q_{xz},Q_{yz})$ to the in-plane field, and is therefore a pseudoscalar: it vanishes for any island possessing a mirror plane or an inversion center---including the untwisted AA and AB bilayers---and reverses sign under $\theta\to-\theta$. Since decoupled layers are individually achiral, a finite $Q_{yz}^{(x)}$, and hence a finite $C$, is a direct signature of chirality mediated by interlayer coupling. We report $\Ree\{C\}$, $\Imm\{C\}$, and $|C|$ in Figs.~\ref{fig5} and~\ref{fig6}; by symmetry, $C(-\theta)=-C(\theta)$. Because $C$ is a product of two complex response functions, the signs of its real and imaginary parts depend on the relative phase of $p_x$ and $Q_{yz}$; we therefore use $|C|$ and its reversal under $\theta\to-\theta$ as the chirality diagnostics.

Circular dichroism follows from the same interlayer quadrupole. For layers at $z=\pm b_0/2$ carrying in-plane currents, the induced magnetic dipole $\mb=(1/2c)\int{\rm d}^3\rb\,(\rb\times\Jb^{\rm ind})$ is $m_x=(\ii\ww/2c)Q_{yz}$ and $m_y=-(\ii\ww/2c)Q_{xz}$, so the out-of-phase layer dipoles constitute an in-plane magnetic moment (neglecting interlayer tunneling currents). For light incident along $\zz$, the helicity-dependent absorption is then governed, to leading order in the retardation across the bilayer, by the quadrupole induced by each linearly polarized drive,
\begin{equation} \label{eq:CD}
    \sigma_{\rm CD}(\ww)\propto\frac{\ww^2}{c^2}\Imm\left\{Q_{yz}^{(x)}(\ww)-Q_{xz}^{(y)}(\ww)\right\} ,
\end{equation}
where $\sigma_{\rm CD}\equiv\sigma_{+}-\sigma_{-}$ is the difference in absorption cross section between left- and right-circularly polarized light, and the superscript denotes the drive polarization. The three-fold symmetry of the twisted islands ($C_3\subset D_6,D_3$) imposes $Q_{xz}^{(y)}=-Q_{yz}^{(x)}$, so $\sigma_{\rm CD}\propto(\ww/c)^2 \Imm\{Q_{yz}^{(x)}\}$ follows from the $x$-polarized simulation alone. Note that $\sigma_{\rm CD}$ is linear in $Q_{yz}^{(x)}$, whereas $C$ also contains the resonant dipole $p_x$; $C$ thus quantifies the handedness of the induced charge distribution rather than the circular-dichroism spectrum itself.

\begin{acknowledgments}
The Center for Polariton-driven Light--Matter Interactions (POLIMA) is funded by the Danish National Research Foundation (Project No.~DNRF165).
E.~J.~C.~D. acknowledges support from the European Union (Marie Sk\l{}odowska-Curie Actions, grant agreement No. 101209876).
Part of the computational work was performed on the DeiC Interactive HPC system managed by the eScience Center at the University of Southern Denmark.
\end{acknowledgments}

\section*{Author Contributions}

E.~J.~C.~D. and J.~D.~C. conceived and supervised the study. M.~H. performed the calculations, with input and analysis from all authors. All authors contributed to the development of the theory and writing of the manuscript.

\section*{Data availability}

All data needed to evaluate the conclusions in the paper are present in the paper. Additional data related to this paper may be requested from the authors.

\section*{Competing Interests}

The authors declare no competing interests.


\begin{thebibliography}{42}%
\makeatletter
\providecommand \@ifxundefined [1]{%
 \@ifx{#1\undefined}
}%
\providecommand \@ifnum [1]{%
 \ifnum #1\expandafter \@firstoftwo
 \else \expandafter \@secondoftwo
 \fi
}%
\providecommand \@ifx [1]{%
 \ifx #1\expandafter \@firstoftwo
 \else \expandafter \@secondoftwo
 \fi
}%
\providecommand \natexlab [1]{#1}%
\providecommand \enquote  [1]{``#1''}%
\providecommand \bibnamefont  [1]{#1}%
\providecommand \bibfnamefont [1]{#1}%
\providecommand \citenamefont [1]{#1}%
\providecommand \href@noop [0]{\@secondoftwo}%
\providecommand \href [0]{\begingroup \@sanitize@url \@href}%
\providecommand \@href[1]{\@@startlink{#1}\@@href}%
\providecommand \@@href[1]{\endgroup#1\@@endlink}%
\providecommand \@sanitize@url [0]{\catcode `\\12\catcode `\$12\catcode `\&12\catcode `\#12\catcode `\^12\catcode `\_12\catcode `\%12\relax}%
\providecommand \@@startlink[1]{}%
\providecommand \@@endlink[0]{}%
\providecommand \url  [0]{\begingroup\@sanitize@url \@url }%
\providecommand \@url [1]{\endgroup\@href {#1}{\urlprefix }}%
\providecommand \urlprefix  [0]{URL }%
\providecommand \Eprint [0]{\href }%
\providecommand \doibase [0]{https://doi.org/}%
\providecommand \selectlanguage [0]{\@gobble}%
\providecommand \bibinfo  [0]{\@secondoftwo}%
\providecommand \bibfield  [0]{\@secondoftwo}%
\providecommand \translation [1]{[#1]}%
\providecommand \BibitemOpen [0]{}%
\providecommand \bibitemStop [0]{}%
\providecommand \bibitemNoStop [0]{.\EOS\space}%
\providecommand \EOS [0]{\spacefactor3000\relax}%
\providecommand \BibitemShut  [1]{\csname bibitem#1\endcsname}%
\let\auto@bib@innerbib\@empty
\bibitem [{\citenamefont {Novoselov}\ \emph {et~al.}(2004)\citenamefont {Novoselov}, \citenamefont {Geim}, \citenamefont {Morozov}, \citenamefont {Jiang}, \citenamefont {Zhang}, \citenamefont {Dubonos}, \citenamefont {Grigorieva},\ and\ \citenamefont {Firsov}}]{novoselov2004electric}%
  \BibitemOpen
  \bibfield  {author} {\bibinfo {author} {\bibfnamefont {K.~S.}\ \bibnamefont {Novoselov}}, \bibinfo {author} {\bibfnamefont {A.~K.}\ \bibnamefont {Geim}}, \bibinfo {author} {\bibfnamefont {S.~V.}\ \bibnamefont {Morozov}}, \bibinfo {author} {\bibfnamefont {D.}~\bibnamefont {Jiang}}, \bibinfo {author} {\bibfnamefont {Y.}~\bibnamefont {Zhang}}, \bibinfo {author} {\bibfnamefont {S.~V.}\ \bibnamefont {Dubonos}}, \bibinfo {author} {\bibfnamefont {I.~V.}\ \bibnamefont {Grigorieva}},\ and\ \bibinfo {author} {\bibfnamefont {A.~A.}\ \bibnamefont {Firsov}},\ }\bibfield  {title} {\bibinfo {title} {Electric field effect in atomically thin carbon films},\ }\href {https://doi.org/10.1126/science.1102896} {\bibfield  {journal} {\bibinfo  {journal} {Science}\ }\textbf {\bibinfo {volume} {306}},\ \bibinfo {pages} {666} (\bibinfo {year} {2004})}\BibitemShut {NoStop}%
\bibitem [{\citenamefont {Novoselov}\ \emph {et~al.}(2005)\citenamefont {Novoselov}, \citenamefont {Geim}, \citenamefont {Morozov}, \citenamefont {Jiang}, \citenamefont {Katsnelson}, \citenamefont {Grigorieva}, \citenamefont {Dubonos},\ and\ \citenamefont {Firsov}}]{novoselov2005twodimensional}%
  \BibitemOpen
  \bibfield  {author} {\bibinfo {author} {\bibfnamefont {K.~S.}\ \bibnamefont {Novoselov}}, \bibinfo {author} {\bibfnamefont {A.~K.}\ \bibnamefont {Geim}}, \bibinfo {author} {\bibfnamefont {S.~V.}\ \bibnamefont {Morozov}}, \bibinfo {author} {\bibfnamefont {D.}~\bibnamefont {Jiang}}, \bibinfo {author} {\bibfnamefont {M.~I.}\ \bibnamefont {Katsnelson}}, \bibinfo {author} {\bibfnamefont {I.~V.}\ \bibnamefont {Grigorieva}}, \bibinfo {author} {\bibfnamefont {S.~V.}\ \bibnamefont {Dubonos}},\ and\ \bibinfo {author} {\bibfnamefont {A.~A.}\ \bibnamefont {Firsov}},\ }\bibfield  {title} {\bibinfo {title} {Two-dimensional gas of massless {Dirac} fermions in graphene},\ }\href {https://doi.org/10.1038/nature04233} {\bibfield  {journal} {\bibinfo  {journal} {Nature}\ }\textbf {\bibinfo {volume} {438}},\ \bibinfo {pages} {197} (\bibinfo {year} {2005})}\BibitemShut {NoStop}%
\bibitem [{\citenamefont {Nair}\ \emph {et~al.}(2008)\citenamefont {Nair}, \citenamefont {Blake}, \citenamefont {Grigorenko}, \citenamefont {Novoselov}, \citenamefont {Booth}, \citenamefont {Stauber}, \citenamefont {Peres},\ and\ \citenamefont {Geim}}]{nair2008fine}%
  \BibitemOpen
  \bibfield  {author} {\bibinfo {author} {\bibfnamefont {R.~R.}\ \bibnamefont {Nair}}, \bibinfo {author} {\bibfnamefont {P.}~\bibnamefont {Blake}}, \bibinfo {author} {\bibfnamefont {A.~N.}\ \bibnamefont {Grigorenko}}, \bibinfo {author} {\bibfnamefont {K.~S.}\ \bibnamefont {Novoselov}}, \bibinfo {author} {\bibfnamefont {T.~J.}\ \bibnamefont {Booth}}, \bibinfo {author} {\bibfnamefont {T.}~\bibnamefont {Stauber}}, \bibinfo {author} {\bibfnamefont {N.~M.~R.}\ \bibnamefont {Peres}},\ and\ \bibinfo {author} {\bibfnamefont {A.~K.}\ \bibnamefont {Geim}},\ }\bibfield  {title} {\bibinfo {title} {Fine structure constant defines visual transparency of graphene},\ }\href {https://doi.org/10.1126/science.1156965} {\bibfield  {journal} {\bibinfo  {journal} {Science}\ }\textbf {\bibinfo {volume} {320}},\ \bibinfo {pages} {1308} (\bibinfo {year} {2008})}\BibitemShut {NoStop}%
\bibitem [{\citenamefont {Bonaccorso}\ \emph {et~al.}(2010)\citenamefont {Bonaccorso}, \citenamefont {Sun}, \citenamefont {Hasan},\ and\ \citenamefont {Ferrari}}]{bonaccorso2010graphene}%
  \BibitemOpen
  \bibfield  {author} {\bibinfo {author} {\bibfnamefont {F.}~\bibnamefont {Bonaccorso}}, \bibinfo {author} {\bibfnamefont {Z.}~\bibnamefont {Sun}}, \bibinfo {author} {\bibfnamefont {T.}~\bibnamefont {Hasan}},\ and\ \bibinfo {author} {\bibfnamefont {A.~C.}\ \bibnamefont {Ferrari}},\ }\bibfield  {title} {\bibinfo {title} {Graphene photonics and optoelectronics},\ }\href {https://doi.org/10.1038/nphoton.2010.186} {\bibfield  {journal} {\bibinfo  {journal} {Nat.\ Photonics}\ }\textbf {\bibinfo {volume} {4}},\ \bibinfo {pages} {611} (\bibinfo {year} {2010})}\BibitemShut {NoStop}%
\bibitem [{\citenamefont {Cox}\ \emph {et~al.}(2016)\citenamefont {Cox}, \citenamefont {Silveiro},\ and\ \citenamefont {{Garc{\'\i}a de Abajo}}}]{cox2016quantum}%
  \BibitemOpen
  \bibfield  {author} {\bibinfo {author} {\bibfnamefont {J.~D.}\ \bibnamefont {Cox}}, \bibinfo {author} {\bibfnamefont {I.}~\bibnamefont {Silveiro}},\ and\ \bibinfo {author} {\bibfnamefont {F.~J.}\ \bibnamefont {{Garc{\'\i}a de Abajo}}},\ }\bibfield  {title} {\bibinfo {title} {Quantum effects in the nonlinear response of graphene plasmons},\ }\href {https://doi.org/10.1021/acsnano.5b06110} {\bibfield  {journal} {\bibinfo  {journal} {ACS Nano}\ }\textbf {\bibinfo {volume} {10}},\ \bibinfo {pages} {1995} (\bibinfo {year} {2016})}\BibitemShut {NoStop}%
\bibitem [{\citenamefont {{Lopes dos Santos}}\ \emph {et~al.}(2007)\citenamefont {{Lopes dos Santos}}, \citenamefont {Peres},\ and\ \citenamefont {{Castro Neto}}}]{lopesdossantos2007graphene}%
  \BibitemOpen
  \bibfield  {author} {\bibinfo {author} {\bibfnamefont {J.~M.~B.}\ \bibnamefont {{Lopes dos Santos}}}, \bibinfo {author} {\bibfnamefont {N.~M.~R.}\ \bibnamefont {Peres}},\ and\ \bibinfo {author} {\bibfnamefont {A.~H.}\ \bibnamefont {{Castro Neto}}},\ }\bibfield  {title} {\bibinfo {title} {Graphene bilayer with a twist: Electronic structure},\ }\href {https://doi.org/10.1103/PhysRevLett.99.256802} {\bibfield  {journal} {\bibinfo  {journal} {Phys.\ Rev.\ Lett.}\ }\textbf {\bibinfo {volume} {99}},\ \bibinfo {pages} {256802} (\bibinfo {year} {2007})}\BibitemShut {NoStop}%
\bibitem [{\citenamefont {Cao}\ \emph {et~al.}(2018{\natexlab{a}})\citenamefont {Cao}, \citenamefont {Fatemi}, \citenamefont {Fang}, \citenamefont {Watanabe}, \citenamefont {Taniguchi}, \citenamefont {Kaxiras},\ and\ \citenamefont {Jarillo-Herrero}}]{cao2018unconventional}%
  \BibitemOpen
  \bibfield  {author} {\bibinfo {author} {\bibfnamefont {Y.}~\bibnamefont {Cao}}, \bibinfo {author} {\bibfnamefont {V.}~\bibnamefont {Fatemi}}, \bibinfo {author} {\bibfnamefont {S.}~\bibnamefont {Fang}}, \bibinfo {author} {\bibfnamefont {K.}~\bibnamefont {Watanabe}}, \bibinfo {author} {\bibfnamefont {T.}~\bibnamefont {Taniguchi}}, \bibinfo {author} {\bibfnamefont {E.}~\bibnamefont {Kaxiras}},\ and\ \bibinfo {author} {\bibfnamefont {P.}~\bibnamefont {Jarillo-Herrero}},\ }\bibfield  {title} {\bibinfo {title} {Unconventional superconductivity in magic-angle graphene superlattices},\ }\href {https://doi.org/10.1038/nature26160} {\bibfield  {journal} {\bibinfo  {journal} {Nature}\ }\textbf {\bibinfo {volume} {556}},\ \bibinfo {pages} {43} (\bibinfo {year} {2018}{\natexlab{a}})}\BibitemShut {NoStop}%
\bibitem [{\citenamefont {Andrei}\ and\ \citenamefont {MacDonald}(2020)}]{andrei2020graphene}%
  \BibitemOpen
  \bibfield  {author} {\bibinfo {author} {\bibfnamefont {E.~Y.}\ \bibnamefont {Andrei}}\ and\ \bibinfo {author} {\bibfnamefont {A.~H.}\ \bibnamefont {MacDonald}},\ }\bibfield  {title} {\bibinfo {title} {Graphene bilayers with a twist},\ }\href {https://doi.org/10.1038/s41563-020-00840-0} {\bibfield  {journal} {\bibinfo  {journal} {Nat.\ Mater.}\ }\textbf {\bibinfo {volume} {19}},\ \bibinfo {pages} {1265} (\bibinfo {year} {2020})}\BibitemShut {NoStop}%
\bibitem [{\citenamefont {Koshino}\ and\ \citenamefont {Nam}(2020)}]{koshino2020effective}%
  \BibitemOpen
  \bibfield  {author} {\bibinfo {author} {\bibfnamefont {M.}~\bibnamefont {Koshino}}\ and\ \bibinfo {author} {\bibfnamefont {N.~N.~T.}\ \bibnamefont {Nam}},\ }\bibfield  {title} {\bibinfo {title} {Effective continuum model for relaxed twisted bilayer graphene and moir{\'e} electron-phonon interaction},\ }\href {https://doi.org/10.1103/PhysRevB.101.195425} {\bibfield  {journal} {\bibinfo  {journal} {Phys.\ Rev.\ B}\ }\textbf {\bibinfo {volume} {101}},\ \bibinfo {pages} {195425} (\bibinfo {year} {2020})}\BibitemShut {NoStop}%
\bibitem [{\citenamefont {Dey}\ \emph {et~al.}(2023)\citenamefont {Dey}, \citenamefont {Chowdhury}, \citenamefont {Pe{\~n}a}, \citenamefont {Singh}, \citenamefont {Wu},\ and\ \citenamefont {Askari}}]{dey2023atomistic}%
  \BibitemOpen
  \bibfield  {author} {\bibinfo {author} {\bibfnamefont {A.}~\bibnamefont {Dey}}, \bibinfo {author} {\bibfnamefont {S.~A.}\ \bibnamefont {Chowdhury}}, \bibinfo {author} {\bibfnamefont {T.}~\bibnamefont {Pe{\~n}a}}, \bibinfo {author} {\bibfnamefont {S.}~\bibnamefont {Singh}}, \bibinfo {author} {\bibfnamefont {S.~M.}\ \bibnamefont {Wu}},\ and\ \bibinfo {author} {\bibfnamefont {H.}~\bibnamefont {Askari}},\ }\bibfield  {title} {\bibinfo {title} {An atomistic insight into moir{\'e} reconstruction in twisted bilayer graphene beyond the magic angle},\ }\href {https://doi.org/10.1021/acsaenm.2c00259} {\bibfield  {journal} {\bibinfo  {journal} {ACS Appl.\ Eng.\ Mater.}\ }\textbf {\bibinfo {volume} {1}},\ \bibinfo {pages} {970} (\bibinfo {year} {2023})}\BibitemShut {NoStop}%
\bibitem [{\citenamefont {Bistritzer}\ and\ \citenamefont {MacDonald}(2011)}]{bistritzer2011moire}%
  \BibitemOpen
  \bibfield  {author} {\bibinfo {author} {\bibfnamefont {R.}~\bibnamefont {Bistritzer}}\ and\ \bibinfo {author} {\bibfnamefont {A.~H.}\ \bibnamefont {MacDonald}},\ }\bibfield  {title} {\bibinfo {title} {Moir{\'e} bands in twisted double-layer graphene},\ }\href {https://doi.org/10.1073/pnas.1108174108} {\bibfield  {journal} {\bibinfo  {journal} {Proc.\ Natl.\ Acad.\ Sci.\ U.S.A.}\ }\textbf {\bibinfo {volume} {108}},\ \bibinfo {pages} {12233} (\bibinfo {year} {2011})}\BibitemShut {NoStop}%
\bibitem [{\citenamefont {{Trambly de Laissardi{\`e}re}}\ \emph {et~al.}(2012)\citenamefont {{Trambly de Laissardi{\`e}re}}, \citenamefont {Mayou},\ and\ \citenamefont {Magaud}}]{tramblydelaissardiere2012numerical}%
  \BibitemOpen
  \bibfield  {author} {\bibinfo {author} {\bibfnamefont {G.}~\bibnamefont {{Trambly de Laissardi{\`e}re}}}, \bibinfo {author} {\bibfnamefont {D.}~\bibnamefont {Mayou}},\ and\ \bibinfo {author} {\bibfnamefont {L.}~\bibnamefont {Magaud}},\ }\bibfield  {title} {\bibinfo {title} {Numerical studies of confined states in rotated bilayers of graphene},\ }\href {https://doi.org/10.1103/PhysRevB.86.125413} {\bibfield  {journal} {\bibinfo  {journal} {Phys.\ Rev.\ B}\ }\textbf {\bibinfo {volume} {86}},\ \bibinfo {pages} {125413} (\bibinfo {year} {2012})}\BibitemShut {NoStop}%
\bibitem [{\citenamefont {Cao}\ \emph {et~al.}(2018{\natexlab{b}})\citenamefont {Cao}, \citenamefont {Fatemi}, \citenamefont {Demir}, \citenamefont {Fang}, \citenamefont {Tomarken}, \citenamefont {Luo}, \citenamefont {Sanchez-Yamagishi}, \citenamefont {Watanabe}, \citenamefont {Taniguchi}, \citenamefont {Kaxiras}, \citenamefont {Ashoori},\ and\ \citenamefont {Jarillo-Herrero}}]{cao2018correlated}%
  \BibitemOpen
  \bibfield  {author} {\bibinfo {author} {\bibfnamefont {Y.}~\bibnamefont {Cao}}, \bibinfo {author} {\bibfnamefont {V.}~\bibnamefont {Fatemi}}, \bibinfo {author} {\bibfnamefont {A.}~\bibnamefont {Demir}}, \bibinfo {author} {\bibfnamefont {S.}~\bibnamefont {Fang}}, \bibinfo {author} {\bibfnamefont {S.~L.}\ \bibnamefont {Tomarken}}, \bibinfo {author} {\bibfnamefont {J.~Y.}\ \bibnamefont {Luo}}, \bibinfo {author} {\bibfnamefont {J.~D.}\ \bibnamefont {Sanchez-Yamagishi}}, \bibinfo {author} {\bibfnamefont {K.}~\bibnamefont {Watanabe}}, \bibinfo {author} {\bibfnamefont {T.}~\bibnamefont {Taniguchi}}, \bibinfo {author} {\bibfnamefont {E.}~\bibnamefont {Kaxiras}}, \bibinfo {author} {\bibfnamefont {R.~C.}\ \bibnamefont {Ashoori}},\ and\ \bibinfo {author} {\bibfnamefont {P.}~\bibnamefont {Jarillo-Herrero}},\ }\bibfield  {title} {\bibinfo {title} {Correlated insulator behaviour at half-filling in magic-angle graphene superlattices},\ }\href {https://doi.org/10.1038/nature26154} {\bibfield  {journal} {\bibinfo  {journal}
  {Nature}\ }\textbf {\bibinfo {volume} {556}},\ \bibinfo {pages} {80} (\bibinfo {year} {2018}{\natexlab{b}})}\BibitemShut {NoStop}%
\bibitem [{\citenamefont {Kuang}\ \emph {et~al.}(2025)\citenamefont {Kuang}, \citenamefont {Escudero}, \citenamefont {Pantale{\'o}n}, \citenamefont {Guinea},\ and\ \citenamefont {Zhan}}]{kuang2025review}%
  \BibitemOpen
  \bibfield  {author} {\bibinfo {author} {\bibfnamefont {X.}~\bibnamefont {Kuang}}, \bibinfo {author} {\bibfnamefont {F.}~\bibnamefont {Escudero}}, \bibinfo {author} {\bibfnamefont {P.~A.}\ \bibnamefont {Pantale{\'o}n}}, \bibinfo {author} {\bibfnamefont {F.}~\bibnamefont {Guinea}},\ and\ \bibinfo {author} {\bibfnamefont {Z.}~\bibnamefont {Zhan}},\ }\bibfield  {title} {\bibinfo {title} {Review of the tight-binding method applicable to the properties of moir{\'e} superlattices},\ }\href {https://doi.org/10.1039/d5cp03472h} {\bibfield  {journal} {\bibinfo  {journal} {Phys.\ Chem.\ Chem.\ Phys.}\ }\textbf {\bibinfo {volume} {27}},\ \bibinfo {pages} {25232} (\bibinfo {year} {2025})}\BibitemShut {NoStop}%
\bibitem [{\citenamefont {{Garc{\'\i}a de Abajo}}(2014)}]{garciadeabajo2014graphene}%
  \BibitemOpen
  \bibfield  {author} {\bibinfo {author} {\bibfnamefont {F.~J.}\ \bibnamefont {{Garc{\'\i}a de Abajo}}},\ }\bibfield  {title} {\bibinfo {title} {Graphene plasmonics: challenges and opportunities},\ }\href {https://doi.org/10.1021/ph400147y} {\bibfield  {journal} {\bibinfo  {journal} {ACS Photonics}\ }\textbf {\bibinfo {volume} {1}},\ \bibinfo {pages} {135} (\bibinfo {year} {2014})}\BibitemShut {NoStop}%
\bibitem [{\citenamefont {Thongrattanasiri}\ \emph {et~al.}(2012)\citenamefont {Thongrattanasiri}, \citenamefont {Manjavacas},\ and\ \citenamefont {{Garc{\'\i}a de Abajo}}}]{thongrattanasiri2012quantum}%
  \BibitemOpen
  \bibfield  {author} {\bibinfo {author} {\bibfnamefont {S.}~\bibnamefont {Thongrattanasiri}}, \bibinfo {author} {\bibfnamefont {A.}~\bibnamefont {Manjavacas}},\ and\ \bibinfo {author} {\bibfnamefont {F.~J.}\ \bibnamefont {{Garc{\'\i}a de Abajo}}},\ }\bibfield  {title} {\bibinfo {title} {Quantum finite-size effects in graphene plasmons},\ }\href {https://doi.org/10.1021/nn204780e} {\bibfield  {journal} {\bibinfo  {journal} {ACS Nano}\ }\textbf {\bibinfo {volume} {6}},\ \bibinfo {pages} {1766} (\bibinfo {year} {2012})}\BibitemShut {NoStop}%
\bibitem [{\citenamefont {Cox}\ and\ \citenamefont {{Garc{\'\i}a de Abajo}}(2014)}]{cox2014electrically}%
  \BibitemOpen
  \bibfield  {author} {\bibinfo {author} {\bibfnamefont {J.~D.}\ \bibnamefont {Cox}}\ and\ \bibinfo {author} {\bibfnamefont {F.~J.}\ \bibnamefont {{Garc{\'\i}a de Abajo}}},\ }\bibfield  {title} {\bibinfo {title} {Electrically tunable nonlinear plasmonics in graphene nanoislands},\ }\href {https://doi.org/10.1038/ncomms6725} {\bibfield  {journal} {\bibinfo  {journal} {Nat.\ Commun.}\ }\textbf {\bibinfo {volume} {5}},\ \bibinfo {pages} {5725} (\bibinfo {year} {2014})}\BibitemShut {NoStop}%
\bibitem [{\citenamefont {Silveiro}\ \emph {et~al.}(2015)\citenamefont {Silveiro}, \citenamefont {{Plaza Ortega}},\ and\ \citenamefont {{Garc{\'\i}a de Abajo}}}]{silveiro2015quantum}%
  \BibitemOpen
  \bibfield  {author} {\bibinfo {author} {\bibfnamefont {I.}~\bibnamefont {Silveiro}}, \bibinfo {author} {\bibfnamefont {J.~M.}\ \bibnamefont {{Plaza Ortega}}},\ and\ \bibinfo {author} {\bibfnamefont {F.~J.}\ \bibnamefont {{Garc{\'\i}a de Abajo}}},\ }\bibfield  {title} {\bibinfo {title} {Quantum nonlocal effects in individual and interacting graphene nanoribbons},\ }\href {https://doi.org/10.1038/lsa.2015.14} {\bibfield  {journal} {\bibinfo  {journal} {Light Sci.\ Appl.}\ }\textbf {\bibinfo {volume} {4}},\ \bibinfo {pages} {e241} (\bibinfo {year} {2015})}\BibitemShut {NoStop}%
\bibitem [{\citenamefont {{Garc{\'\i}a de Abajo}}\ \emph {et~al.}(2025)\citenamefont {{Garc{\'\i}a de Abajo}}, \citenamefont {Basov}, \citenamefont {Koppens}, \citenamefont {Orsini}, \citenamefont {Ceccanti}, \citenamefont {Castilla}, \citenamefont {Cavicchi}, \citenamefont {Polini}, \citenamefont {Gon{\c{c}}alves}, \citenamefont {Costa}, \citenamefont {Peres}, \citenamefont {Mortensen}, \citenamefont {Bharadwaj}, \citenamefont {Jacob}, \citenamefont {Schuck}, \citenamefont {Pasupathy}, \citenamefont {Delor}, \citenamefont {Liu}, \citenamefont {Mugarza}, \citenamefont {Merino}, \citenamefont {Cuxart}, \citenamefont {Ch{\'a}vez-Angel}, \citenamefont {{\v{S}}vec}, \citenamefont {Tizei}, \citenamefont {Dirnberger}, \citenamefont {Deng}, \citenamefont {Schneider}, \citenamefont {Menon}, \citenamefont {Deilmann}, \citenamefont {Chernikov}, \citenamefont {Thygesen}, \citenamefont {Abate}, \citenamefont {Terrones}, \citenamefont {Sangwan}, \citenamefont {Hersam}, \citenamefont {Yu}, \citenamefont {Chen},
  \citenamefont {Heinz}, \citenamefont {Murthy}, \citenamefont {Kroner}, \citenamefont {Smole{\'n}ski}, \citenamefont {Thureja}, \citenamefont {Chervy}, \citenamefont {Genco}, \citenamefont {Trovatello}, \citenamefont {Cerullo}, \citenamefont {{Dal Conte}}, \citenamefont {Timmer}, \citenamefont {{De Sio}}, \citenamefont {Lienau}, \citenamefont {Shang}, \citenamefont {Hong}, \citenamefont {Liu}, \citenamefont {Sun}, \citenamefont {Rozema}, \citenamefont {Walther}, \citenamefont {Al{\`u}}, \citenamefont {Cotrufo}, \citenamefont {Queiroz}, \citenamefont {Zhu}, \citenamefont {Cox}, \citenamefont {Dias}, \citenamefont {{Rodr{\'\i}guez Echarri}}, \citenamefont {Iyikanat}, \citenamefont {Marini}, \citenamefont {Herrmann}, \citenamefont {Tornow}, \citenamefont {Klimmer}, \citenamefont {Wilhelm}, \citenamefont {Soavi}, \citenamefont {Sun}, \citenamefont {Wu}, \citenamefont {Xiong}, \citenamefont {Matsyshyn}, \citenamefont {{Krishna Kumar}}, \citenamefont {Song}, \citenamefont {Bucher}, \citenamefont {Gorlach},
  \citenamefont {Tsesses}, \citenamefont {Kaminer}, \citenamefont {Schwab}, \citenamefont {Mangold}, \citenamefont {Giessen}, \citenamefont {{S{\'a}nchez S{\'a}nchez}}, \citenamefont {Efetov}, \citenamefont {Low}, \citenamefont {G{\'o}mez-Santos}, \citenamefont {Stauber}, \citenamefont {{\'A}lvarez-P{\'e}rez}, \citenamefont {Duan}, \citenamefont {Mart{\'\i}n-Moreno}, \citenamefont {Paarmann}, \citenamefont {Caldwell}, \citenamefont {Nikitin}, \citenamefont {Alonso-Gonz{\'a}lez}, \citenamefont {Mueller}, \citenamefont {Volkov}, \citenamefont {Jariwala}, \citenamefont {Shegai}, \citenamefont {van~de Groep}, \citenamefont {Boltasseva}, \citenamefont {Bondarev}, \citenamefont {Shalaev}, \citenamefont {Simon}, \citenamefont {Fruhling}, \citenamefont {Shen}, \citenamefont {Novko}, \citenamefont {Tan}, \citenamefont {Wang}, \citenamefont {Petek}, \citenamefont {Mkhitaryan}, \citenamefont {Yu}, \citenamefont {Manjavacas}, \citenamefont {Ortega}, \citenamefont {Cheng}, \citenamefont {Tian}, \citenamefont {Mao},
  \citenamefont {{Van Thourhout}}, \citenamefont {Gan}, \citenamefont {Dai}, \citenamefont {Sternbach}, \citenamefont {Zhou}, \citenamefont {Hafezi}, \citenamefont {Litvinov}, \citenamefont {Grzeszczyk}, \citenamefont {Novoselov}, \citenamefont {Koperski}, \citenamefont {Papadopoulos}, \citenamefont {Novotny}, \citenamefont {Viti}, \citenamefont {Vitiello}, \citenamefont {Cottam}, \citenamefont {Dewes}, \citenamefont {Makarovsky}, \citenamefont {Patan{\`e}}, \citenamefont {Song}, \citenamefont {Cai}, \citenamefont {Chen}, \citenamefont {Naveh}, \citenamefont {Jang}, \citenamefont {Park}, \citenamefont {Xia}, \citenamefont {Jenke}, \citenamefont {Bajo}, \citenamefont {Braun}, \citenamefont {Burch}, \citenamefont {Zhao},\ and\ \citenamefont {Xu}}]{garciadeabajo2025roadmap}%
  \BibitemOpen
  \bibfield  {author} {\bibinfo {author} {\bibfnamefont {F.~J.}\ \bibnamefont {{Garc{\'\i}a de Abajo}}}, \bibinfo {author} {\bibfnamefont {D.~N.}\ \bibnamefont {Basov}}, \bibinfo {author} {\bibfnamefont {F.~H.~L.}\ \bibnamefont {Koppens}}, \bibinfo {author} {\bibfnamefont {L.}~\bibnamefont {Orsini}}, \bibinfo {author} {\bibfnamefont {M.}~\bibnamefont {Ceccanti}}, \bibinfo {author} {\bibfnamefont {S.}~\bibnamefont {Castilla}}, \bibinfo {author} {\bibfnamefont {L.}~\bibnamefont {Cavicchi}}, \bibinfo {author} {\bibfnamefont {M.}~\bibnamefont {Polini}}, \bibinfo {author} {\bibfnamefont {P.~A.~D.}\ \bibnamefont {Gon{\c{c}}alves}}, \bibinfo {author} {\bibfnamefont {A.~T.}\ \bibnamefont {Costa}}, \bibinfo {author} {\bibfnamefont {N.~M.~R.}\ \bibnamefont {Peres}}, \bibinfo {author} {\bibfnamefont {N.~A.}\ \bibnamefont {Mortensen}}, \bibinfo {author} {\bibfnamefont {S.}~\bibnamefont {Bharadwaj}}, \bibinfo {author} {\bibfnamefont {Z.}~\bibnamefont {Jacob}}, \bibinfo {author} {\bibfnamefont {P.~J.}\ \bibnamefont
  {Schuck}}, \bibinfo {author} {\bibfnamefont {A.~N.}\ \bibnamefont {Pasupathy}}, \bibinfo {author} {\bibfnamefont {M.}~\bibnamefont {Delor}}, \bibinfo {author} {\bibfnamefont {M.~K.}\ \bibnamefont {Liu}}, \bibinfo {author} {\bibfnamefont {A.}~\bibnamefont {Mugarza}}, \bibinfo {author} {\bibfnamefont {P.}~\bibnamefont {Merino}}, \bibinfo {author} {\bibfnamefont {M.~G.}\ \bibnamefont {Cuxart}}, \bibinfo {author} {\bibfnamefont {E.}~\bibnamefont {Ch{\'a}vez-Angel}}, \bibinfo {author} {\bibfnamefont {M.}~\bibnamefont {{\v{S}}vec}}, \bibinfo {author} {\bibfnamefont {L.~H.~G.}\ \bibnamefont {Tizei}}, \bibinfo {author} {\bibfnamefont {F.}~\bibnamefont {Dirnberger}}, \bibinfo {author} {\bibfnamefont {H.}~\bibnamefont {Deng}}, \bibinfo {author} {\bibfnamefont {C.}~\bibnamefont {Schneider}}, \bibinfo {author} {\bibfnamefont {V.}~\bibnamefont {Menon}}, \bibinfo {author} {\bibfnamefont {T.}~\bibnamefont {Deilmann}}, \bibinfo {author} {\bibfnamefont {A.}~\bibnamefont {Chernikov}}, \bibinfo {author} {\bibfnamefont
  {K.~S.}\ \bibnamefont {Thygesen}}, \bibinfo {author} {\bibfnamefont {Y.}~\bibnamefont {Abate}}, \bibinfo {author} {\bibfnamefont {M.}~\bibnamefont {Terrones}}, \bibinfo {author} {\bibfnamefont {V.~K.}\ \bibnamefont {Sangwan}}, \bibinfo {author} {\bibfnamefont {M.~C.}\ \bibnamefont {Hersam}}, \bibinfo {author} {\bibfnamefont {L.}~\bibnamefont {Yu}}, \bibinfo {author} {\bibfnamefont {X.}~\bibnamefont {Chen}}, \bibinfo {author} {\bibfnamefont {T.~F.}\ \bibnamefont {Heinz}}, \bibinfo {author} {\bibfnamefont {P.}~\bibnamefont {Murthy}}, \bibinfo {author} {\bibfnamefont {M.}~\bibnamefont {Kroner}}, \bibinfo {author} {\bibfnamefont {T.}~\bibnamefont {Smole{\'n}ski}}, \bibinfo {author} {\bibfnamefont {D.}~\bibnamefont {Thureja}}, \bibinfo {author} {\bibfnamefont {T.}~\bibnamefont {Chervy}}, \bibinfo {author} {\bibfnamefont {A.}~\bibnamefont {Genco}}, \bibinfo {author} {\bibfnamefont {C.}~\bibnamefont {Trovatello}}, \bibinfo {author} {\bibfnamefont {G.}~\bibnamefont {Cerullo}}, \bibinfo {author} {\bibfnamefont
  {S.}~\bibnamefont {{Dal Conte}}}, \bibinfo {author} {\bibfnamefont {D.}~\bibnamefont {Timmer}}, \bibinfo {author} {\bibfnamefont {A.}~\bibnamefont {{De Sio}}}, \bibinfo {author} {\bibfnamefont {C.}~\bibnamefont {Lienau}}, \bibinfo {author} {\bibfnamefont {N.}~\bibnamefont {Shang}}, \bibinfo {author} {\bibfnamefont {H.}~\bibnamefont {Hong}}, \bibinfo {author} {\bibfnamefont {K.}~\bibnamefont {Liu}}, \bibinfo {author} {\bibfnamefont {Z.}~\bibnamefont {Sun}}, \bibinfo {author} {\bibfnamefont {L.~A.}\ \bibnamefont {Rozema}}, \bibinfo {author} {\bibfnamefont {P.}~\bibnamefont {Walther}}, \bibinfo {author} {\bibfnamefont {A.}~\bibnamefont {Al{\`u}}}, \bibinfo {author} {\bibfnamefont {M.}~\bibnamefont {Cotrufo}}, \bibinfo {author} {\bibfnamefont {R.}~\bibnamefont {Queiroz}}, \bibinfo {author} {\bibfnamefont {X.-Y.}\ \bibnamefont {Zhu}}, \bibinfo {author} {\bibfnamefont {J.~D.}\ \bibnamefont {Cox}}, \bibinfo {author} {\bibfnamefont {E.~J.~C.}\ \bibnamefont {Dias}}, \bibinfo {author} {\bibfnamefont
  {{\'A}.}~\bibnamefont {{Rodr{\'\i}guez Echarri}}}, \bibinfo {author} {\bibfnamefont {F.}~\bibnamefont {Iyikanat}}, \bibinfo {author} {\bibfnamefont {A.}~\bibnamefont {Marini}}, \bibinfo {author} {\bibfnamefont {P.}~\bibnamefont {Herrmann}}, \bibinfo {author} {\bibfnamefont {N.}~\bibnamefont {Tornow}}, \bibinfo {author} {\bibfnamefont {S.}~\bibnamefont {Klimmer}}, \bibinfo {author} {\bibfnamefont {J.}~\bibnamefont {Wilhelm}}, \bibinfo {author} {\bibfnamefont {G.}~\bibnamefont {Soavi}}, \bibinfo {author} {\bibfnamefont {Z.}~\bibnamefont {Sun}}, \bibinfo {author} {\bibfnamefont {S.}~\bibnamefont {Wu}}, \bibinfo {author} {\bibfnamefont {Y.}~\bibnamefont {Xiong}}, \bibinfo {author} {\bibfnamefont {O.}~\bibnamefont {Matsyshyn}}, \bibinfo {author} {\bibfnamefont {R.}~\bibnamefont {{Krishna Kumar}}}, \bibinfo {author} {\bibfnamefont {J.~C.~W.}\ \bibnamefont {Song}}, \bibinfo {author} {\bibfnamefont {T.}~\bibnamefont {Bucher}}, \bibinfo {author} {\bibfnamefont {A.}~\bibnamefont {Gorlach}}, \bibinfo {author}
  {\bibfnamefont {S.}~\bibnamefont {Tsesses}}, \bibinfo {author} {\bibfnamefont {I.}~\bibnamefont {Kaminer}}, \bibinfo {author} {\bibfnamefont {J.}~\bibnamefont {Schwab}}, \bibinfo {author} {\bibfnamefont {F.}~\bibnamefont {Mangold}}, \bibinfo {author} {\bibfnamefont {H.}~\bibnamefont {Giessen}}, \bibinfo {author} {\bibfnamefont {M.}~\bibnamefont {{S{\'a}nchez S{\'a}nchez}}}, \bibinfo {author} {\bibfnamefont {D.~K.}\ \bibnamefont {Efetov}}, \bibinfo {author} {\bibfnamefont {T.}~\bibnamefont {Low}}, \bibinfo {author} {\bibfnamefont {G.}~\bibnamefont {G{\'o}mez-Santos}}, \bibinfo {author} {\bibfnamefont {T.}~\bibnamefont {Stauber}}, \bibinfo {author} {\bibfnamefont {G.}~\bibnamefont {{\'A}lvarez-P{\'e}rez}}, \bibinfo {author} {\bibfnamefont {J.}~\bibnamefont {Duan}}, \bibinfo {author} {\bibfnamefont {L.}~\bibnamefont {Mart{\'\i}n-Moreno}}, \bibinfo {author} {\bibfnamefont {A.}~\bibnamefont {Paarmann}}, \bibinfo {author} {\bibfnamefont {J.~D.}\ \bibnamefont {Caldwell}}, \bibinfo {author} {\bibfnamefont {A.~Y.}\
  \bibnamefont {Nikitin}}, \bibinfo {author} {\bibfnamefont {P.}~\bibnamefont {Alonso-Gonz{\'a}lez}}, \bibinfo {author} {\bibfnamefont {N.~S.}\ \bibnamefont {Mueller}}, \bibinfo {author} {\bibfnamefont {V.}~\bibnamefont {Volkov}}, \bibinfo {author} {\bibfnamefont {D.}~\bibnamefont {Jariwala}}, \bibinfo {author} {\bibfnamefont {T.}~\bibnamefont {Shegai}}, \bibinfo {author} {\bibfnamefont {J.}~\bibnamefont {van~de Groep}}, \bibinfo {author} {\bibfnamefont {A.}~\bibnamefont {Boltasseva}}, \bibinfo {author} {\bibfnamefont {I.~V.}\ \bibnamefont {Bondarev}}, \bibinfo {author} {\bibfnamefont {V.~M.}\ \bibnamefont {Shalaev}}, \bibinfo {author} {\bibfnamefont {J.}~\bibnamefont {Simon}}, \bibinfo {author} {\bibfnamefont {C.}~\bibnamefont {Fruhling}}, \bibinfo {author} {\bibfnamefont {G.}~\bibnamefont {Shen}}, \bibinfo {author} {\bibfnamefont {D.}~\bibnamefont {Novko}}, \bibinfo {author} {\bibfnamefont {S.}~\bibnamefont {Tan}}, \bibinfo {author} {\bibfnamefont {B.}~\bibnamefont {Wang}}, \bibinfo {author} {\bibfnamefont
  {H.}~\bibnamefont {Petek}}, \bibinfo {author} {\bibfnamefont {V.}~\bibnamefont {Mkhitaryan}}, \bibinfo {author} {\bibfnamefont {R.}~\bibnamefont {Yu}}, \bibinfo {author} {\bibfnamefont {A.}~\bibnamefont {Manjavacas}}, \bibinfo {author} {\bibfnamefont {J.~E.}\ \bibnamefont {Ortega}}, \bibinfo {author} {\bibfnamefont {X.}~\bibnamefont {Cheng}}, \bibinfo {author} {\bibfnamefont {R.}~\bibnamefont {Tian}}, \bibinfo {author} {\bibfnamefont {D.}~\bibnamefont {Mao}}, \bibinfo {author} {\bibfnamefont {D.}~\bibnamefont {{Van Thourhout}}}, \bibinfo {author} {\bibfnamefont {X.}~\bibnamefont {Gan}}, \bibinfo {author} {\bibfnamefont {Q.}~\bibnamefont {Dai}}, \bibinfo {author} {\bibfnamefont {A.}~\bibnamefont {Sternbach}}, \bibinfo {author} {\bibfnamefont {Y.}~\bibnamefont {Zhou}}, \bibinfo {author} {\bibfnamefont {M.}~\bibnamefont {Hafezi}}, \bibinfo {author} {\bibfnamefont {D.}~\bibnamefont {Litvinov}}, \bibinfo {author} {\bibfnamefont {M.}~\bibnamefont {Grzeszczyk}}, \bibinfo {author} {\bibfnamefont {K.~S.}\
  \bibnamefont {Novoselov}}, \bibinfo {author} {\bibfnamefont {M.}~\bibnamefont {Koperski}}, \bibinfo {author} {\bibfnamefont {S.}~\bibnamefont {Papadopoulos}}, \bibinfo {author} {\bibfnamefont {L.}~\bibnamefont {Novotny}}, \bibinfo {author} {\bibfnamefont {L.}~\bibnamefont {Viti}}, \bibinfo {author} {\bibfnamefont {M.~S.}\ \bibnamefont {Vitiello}}, \bibinfo {author} {\bibfnamefont {N.~D.}\ \bibnamefont {Cottam}}, \bibinfo {author} {\bibfnamefont {B.~T.}\ \bibnamefont {Dewes}}, \bibinfo {author} {\bibfnamefont {O.}~\bibnamefont {Makarovsky}}, \bibinfo {author} {\bibfnamefont {A.}~\bibnamefont {Patan{\`e}}}, \bibinfo {author} {\bibfnamefont {Y.}~\bibnamefont {Song}}, \bibinfo {author} {\bibfnamefont {M.}~\bibnamefont {Cai}}, \bibinfo {author} {\bibfnamefont {J.}~\bibnamefont {Chen}}, \bibinfo {author} {\bibfnamefont {D.}~\bibnamefont {Naveh}}, \bibinfo {author} {\bibfnamefont {H.}~\bibnamefont {Jang}}, \bibinfo {author} {\bibfnamefont {S.}~\bibnamefont {Park}}, \bibinfo {author} {\bibfnamefont
  {F.}~\bibnamefont {Xia}}, \bibinfo {author} {\bibfnamefont {P.~K.}\ \bibnamefont {Jenke}}, \bibinfo {author} {\bibfnamefont {J.}~\bibnamefont {Bajo}}, \bibinfo {author} {\bibfnamefont {B.}~\bibnamefont {Braun}}, \bibinfo {author} {\bibfnamefont {K.~S.}\ \bibnamefont {Burch}}, \bibinfo {author} {\bibfnamefont {L.}~\bibnamefont {Zhao}},\ and\ \bibinfo {author} {\bibfnamefont {X.}~\bibnamefont {Xu}},\ }\bibfield  {title} {\bibinfo {title} {Roadmap for photonics with {2D} materials},\ }\href {https://doi.org/10.1021/acsphotonics.5c00353} {\bibfield  {journal} {\bibinfo  {journal} {ACS Photonics}\ }\textbf {\bibinfo {volume} {12}},\ \bibinfo {pages} {3961} (\bibinfo {year} {2025})}\BibitemShut {NoStop}%
\bibitem [{\citenamefont {Stauber}\ and\ \citenamefont {Kohler}(2016)}]{stauber2016quasiflat}%
  \BibitemOpen
  \bibfield  {author} {\bibinfo {author} {\bibfnamefont {T.}~\bibnamefont {Stauber}}\ and\ \bibinfo {author} {\bibfnamefont {H.}~\bibnamefont {Kohler}},\ }\bibfield  {title} {\bibinfo {title} {Quasi-flat plasmonic bands in twisted bilayer graphene},\ }\href {https://doi.org/10.1021/acs.nanolett.6b02587} {\bibfield  {journal} {\bibinfo  {journal} {Nano Lett.}\ }\textbf {\bibinfo {volume} {16}},\ \bibinfo {pages} {6844} (\bibinfo {year} {2016})}\BibitemShut {NoStop}%
\bibitem [{\citenamefont {Sunku}\ \emph {et~al.}(2018)\citenamefont {Sunku}, \citenamefont {Ni}, \citenamefont {Jiang}, \citenamefont {Yoo}, \citenamefont {Sternbach}, \citenamefont {McLeod}, \citenamefont {Stauber}, \citenamefont {Xiong}, \citenamefont {Taniguchi}, \citenamefont {Watanabe}, \citenamefont {Kim}, \citenamefont {Fogler},\ and\ \citenamefont {Basov}}]{sunku2018photonic}%
  \BibitemOpen
  \bibfield  {author} {\bibinfo {author} {\bibfnamefont {S.~S.}\ \bibnamefont {Sunku}}, \bibinfo {author} {\bibfnamefont {G.~X.}\ \bibnamefont {Ni}}, \bibinfo {author} {\bibfnamefont {B.~Y.}\ \bibnamefont {Jiang}}, \bibinfo {author} {\bibfnamefont {H.}~\bibnamefont {Yoo}}, \bibinfo {author} {\bibfnamefont {A.}~\bibnamefont {Sternbach}}, \bibinfo {author} {\bibfnamefont {A.~S.}\ \bibnamefont {McLeod}}, \bibinfo {author} {\bibfnamefont {T.}~\bibnamefont {Stauber}}, \bibinfo {author} {\bibfnamefont {L.}~\bibnamefont {Xiong}}, \bibinfo {author} {\bibfnamefont {T.}~\bibnamefont {Taniguchi}}, \bibinfo {author} {\bibfnamefont {K.}~\bibnamefont {Watanabe}}, \bibinfo {author} {\bibfnamefont {P.}~\bibnamefont {Kim}}, \bibinfo {author} {\bibfnamefont {M.~M.}\ \bibnamefont {Fogler}},\ and\ \bibinfo {author} {\bibfnamefont {D.~N.}\ \bibnamefont {Basov}},\ }\bibfield  {title} {\bibinfo {title} {Photonic crystals for nano-light in moir{\'e} graphene superlattices},\ }\href {https://doi.org/10.1126/science.aau5144}
  {\bibfield  {journal} {\bibinfo  {journal} {Science}\ }\textbf {\bibinfo {volume} {362}},\ \bibinfo {pages} {1153} (\bibinfo {year} {2018})}\BibitemShut {NoStop}%
\bibitem [{\citenamefont {Hesp}\ \emph {et~al.}(2021)\citenamefont {Hesp}, \citenamefont {Torre}, \citenamefont {Rodan-Legrain}, \citenamefont {Novelli}, \citenamefont {Cao}, \citenamefont {Carr}, \citenamefont {Fang}, \citenamefont {Stepanov}, \citenamefont {Barcons-Ruiz}, \citenamefont {{Herzig Sheinfux}}, \citenamefont {Watanabe}, \citenamefont {Taniguchi}, \citenamefont {Efetov}, \citenamefont {Kaxiras}, \citenamefont {Jarillo-Herrero}, \citenamefont {Polini},\ and\ \citenamefont {Koppens}}]{hesp2021observation}%
  \BibitemOpen
  \bibfield  {author} {\bibinfo {author} {\bibfnamefont {N.~C.~H.}\ \bibnamefont {Hesp}}, \bibinfo {author} {\bibfnamefont {I.}~\bibnamefont {Torre}}, \bibinfo {author} {\bibfnamefont {D.}~\bibnamefont {Rodan-Legrain}}, \bibinfo {author} {\bibfnamefont {P.}~\bibnamefont {Novelli}}, \bibinfo {author} {\bibfnamefont {Y.}~\bibnamefont {Cao}}, \bibinfo {author} {\bibfnamefont {S.}~\bibnamefont {Carr}}, \bibinfo {author} {\bibfnamefont {S.}~\bibnamefont {Fang}}, \bibinfo {author} {\bibfnamefont {P.}~\bibnamefont {Stepanov}}, \bibinfo {author} {\bibfnamefont {D.}~\bibnamefont {Barcons-Ruiz}}, \bibinfo {author} {\bibfnamefont {H.}~\bibnamefont {{Herzig Sheinfux}}}, \bibinfo {author} {\bibfnamefont {K.}~\bibnamefont {Watanabe}}, \bibinfo {author} {\bibfnamefont {T.}~\bibnamefont {Taniguchi}}, \bibinfo {author} {\bibfnamefont {D.~K.}\ \bibnamefont {Efetov}}, \bibinfo {author} {\bibfnamefont {E.}~\bibnamefont {Kaxiras}}, \bibinfo {author} {\bibfnamefont {P.}~\bibnamefont {Jarillo-Herrero}}, \bibinfo {author}
  {\bibfnamefont {M.}~\bibnamefont {Polini}},\ and\ \bibinfo {author} {\bibfnamefont {F.~H.~L.}\ \bibnamefont {Koppens}},\ }\bibfield  {title} {\bibinfo {title} {Observation of interband collective excitations in twisted bilayer graphene},\ }\href {https://doi.org/10.1038/s41567-021-01327-8} {\bibfield  {journal} {\bibinfo  {journal} {Nat.\ Phys.}\ }\textbf {\bibinfo {volume} {17}},\ \bibinfo {pages} {1162} (\bibinfo {year} {2021})}\BibitemShut {NoStop}%
\bibitem [{\citenamefont {Moon}\ and\ \citenamefont {Koshino}(2013)}]{moon2013optical}%
  \BibitemOpen
  \bibfield  {author} {\bibinfo {author} {\bibfnamefont {P.}~\bibnamefont {Moon}}\ and\ \bibinfo {author} {\bibfnamefont {M.}~\bibnamefont {Koshino}},\ }\bibfield  {title} {\bibinfo {title} {Optical absorption in twisted bilayer graphene},\ }\href {https://doi.org/10.1103/PhysRevB.87.205404} {\bibfield  {journal} {\bibinfo  {journal} {Phys.\ Rev.\ B}\ }\textbf {\bibinfo {volume} {87}},\ \bibinfo {pages} {205404} (\bibinfo {year} {2013})}\BibitemShut {NoStop}%
\bibitem [{\citenamefont {Tiutiunnyk}\ \emph {et~al.}(2019)\citenamefont {Tiutiunnyk}, \citenamefont {Duque}, \citenamefont {Caro-Lopera}, \citenamefont {Mora-Ramos},\ and\ \citenamefont {Correa}}]{tiutiunnyk2019opto}%
  \BibitemOpen
  \bibfield  {author} {\bibinfo {author} {\bibfnamefont {A.}~\bibnamefont {Tiutiunnyk}}, \bibinfo {author} {\bibfnamefont {C.~A.}\ \bibnamefont {Duque}}, \bibinfo {author} {\bibfnamefont {F.~J.}\ \bibnamefont {Caro-Lopera}}, \bibinfo {author} {\bibfnamefont {M.~E.}\ \bibnamefont {Mora-Ramos}},\ and\ \bibinfo {author} {\bibfnamefont {J.~D.}\ \bibnamefont {Correa}},\ }\bibfield  {title} {\bibinfo {title} {Opto-electronic properties of twisted bilayer graphene quantum dots},\ }\href {https://doi.org/10.1016/j.physe.2019.03.028} {\bibfield  {journal} {\bibinfo  {journal} {Physica E}\ }\textbf {\bibinfo {volume} {112}},\ \bibinfo {pages} {36} (\bibinfo {year} {2019})}\BibitemShut {NoStop}%
\bibitem [{\citenamefont {Tepliakov}\ \emph {et~al.}(2020)\citenamefont {Tepliakov}, \citenamefont {Orlov}, \citenamefont {Kundelev},\ and\ \citenamefont {Rukhlenko}}]{tepliakov2020twisted}%
  \BibitemOpen
  \bibfield  {author} {\bibinfo {author} {\bibfnamefont {N.~V.}\ \bibnamefont {Tepliakov}}, \bibinfo {author} {\bibfnamefont {A.~V.}\ \bibnamefont {Orlov}}, \bibinfo {author} {\bibfnamefont {E.~V.}\ \bibnamefont {Kundelev}},\ and\ \bibinfo {author} {\bibfnamefont {I.~D.}\ \bibnamefont {Rukhlenko}},\ }\bibfield  {title} {\bibinfo {title} {Twisted bilayer graphene quantum dots for chiral nanophotonics},\ }\href {https://doi.org/10.1021/acs.jpcc.0c07416} {\bibfield  {journal} {\bibinfo  {journal} {J.\ Phys.\ Chem.\ C}\ }\textbf {\bibinfo {volume} {124}},\ \bibinfo {pages} {22704} (\bibinfo {year} {2020})}\BibitemShut {NoStop}%
\bibitem [{\citenamefont {Wang}\ \emph {et~al.}(2022)\citenamefont {Wang}, \citenamefont {Yu}, \citenamefont {R{\"o}sner}, \citenamefont {Katsnelson}, \citenamefont {Lin},\ and\ \citenamefont {Yuan}}]{wang2022polarizationdependent}%
  \BibitemOpen
  \bibfield  {author} {\bibinfo {author} {\bibfnamefont {Y.}~\bibnamefont {Wang}}, \bibinfo {author} {\bibfnamefont {G.}~\bibnamefont {Yu}}, \bibinfo {author} {\bibfnamefont {M.}~\bibnamefont {R{\"o}sner}}, \bibinfo {author} {\bibfnamefont {M.~I.}\ \bibnamefont {Katsnelson}}, \bibinfo {author} {\bibfnamefont {H.-Q.}\ \bibnamefont {Lin}},\ and\ \bibinfo {author} {\bibfnamefont {S.}~\bibnamefont {Yuan}},\ }\bibfield  {title} {\bibinfo {title} {Polarization-dependent selection rules and optical spectrum atlas of twisted bilayer graphene quantum dots},\ }\href {https://doi.org/10.1103/PhysRevX.12.021055} {\bibfield  {journal} {\bibinfo  {journal} {Phys.\ Rev.\ X}\ }\textbf {\bibinfo {volume} {12}},\ \bibinfo {pages} {021055} (\bibinfo {year} {2022})}\BibitemShut {NoStop}%
\bibitem [{\citenamefont {Magar}\ and\ \citenamefont {Apalkov}(2026)}]{magar2026ultrafast}%
  \BibitemOpen
  \bibfield  {author} {\bibinfo {author} {\bibfnamefont {K.~R.}\ \bibnamefont {Magar}}\ and\ \bibinfo {author} {\bibfnamefont {V.}~\bibnamefont {Apalkov}},\ }\bibfield  {title} {\bibinfo {title} {Ultrafast nonlinear optical response of twisted bilayer graphene quantum dots},\ }\href {https://doi.org/10.1103/29pd-lbw5} {\bibfield  {journal} {\bibinfo  {journal} {Phys.\ Rev.\ B}\ }\textbf {\bibinfo {volume} {113}},\ \bibinfo {pages} {125420} (\bibinfo {year} {2026})}\BibitemShut {NoStop}%
\bibitem [{\citenamefont {Westerhout}\ \emph {et~al.}(2022)\citenamefont {Westerhout}, \citenamefont {Katsnelson},\ and\ \citenamefont {R{\"o}sner}}]{westerhout2022quantum}%
  \BibitemOpen
  \bibfield  {author} {\bibinfo {author} {\bibfnamefont {T.}~\bibnamefont {Westerhout}}, \bibinfo {author} {\bibfnamefont {M.~I.}\ \bibnamefont {Katsnelson}},\ and\ \bibinfo {author} {\bibfnamefont {M.}~\bibnamefont {R{\"o}sner}},\ }\bibfield  {title} {\bibinfo {title} {Quantum dot-like plasmonic modes in twisted bilayer graphene supercells},\ }\href {https://doi.org/10.1088/2053-1583/ac38ca} {\bibfield  {journal} {\bibinfo  {journal} {2D Mater.}\ }\textbf {\bibinfo {volume} {9}},\ \bibinfo {pages} {014004} (\bibinfo {year} {2022})}\BibitemShut {NoStop}%
\bibitem [{\citenamefont {Stauber}\ \emph {et~al.}(2018)\citenamefont {Stauber}, \citenamefont {Low},\ and\ \citenamefont {G{\'o}mez-Santos}}]{stauber2018chiral}%
  \BibitemOpen
  \bibfield  {author} {\bibinfo {author} {\bibfnamefont {T.}~\bibnamefont {Stauber}}, \bibinfo {author} {\bibfnamefont {T.}~\bibnamefont {Low}},\ and\ \bibinfo {author} {\bibfnamefont {G.}~\bibnamefont {G{\'o}mez-Santos}},\ }\bibfield  {title} {\bibinfo {title} {Chiral response of twisted bilayer graphene},\ }\href {https://doi.org/10.1103/PhysRevLett.120.046801} {\bibfield  {journal} {\bibinfo  {journal} {Phys.\ Rev.\ Lett.}\ }\textbf {\bibinfo {volume} {120}},\ \bibinfo {pages} {046801} (\bibinfo {year} {2018})}\BibitemShut {NoStop}%
\bibitem [{\citenamefont {Huang}\ \emph {et~al.}(2022)\citenamefont {Huang}, \citenamefont {Tu}, \citenamefont {Shen}, \citenamefont {Zheng}, \citenamefont {Wang}, \citenamefont {Wang}, \citenamefont {Khaliji}, \citenamefont {Park}, \citenamefont {Liu}, \citenamefont {Yang}, \citenamefont {Zhang}, \citenamefont {Shao}, \citenamefont {Li}, \citenamefont {Low}, \citenamefont {Shi},\ and\ \citenamefont {Wang}}]{huang2022observation}%
  \BibitemOpen
  \bibfield  {author} {\bibinfo {author} {\bibfnamefont {T.}~\bibnamefont {Huang}}, \bibinfo {author} {\bibfnamefont {X.}~\bibnamefont {Tu}}, \bibinfo {author} {\bibfnamefont {C.}~\bibnamefont {Shen}}, \bibinfo {author} {\bibfnamefont {B.}~\bibnamefont {Zheng}}, \bibinfo {author} {\bibfnamefont {J.}~\bibnamefont {Wang}}, \bibinfo {author} {\bibfnamefont {H.}~\bibnamefont {Wang}}, \bibinfo {author} {\bibfnamefont {K.}~\bibnamefont {Khaliji}}, \bibinfo {author} {\bibfnamefont {S.~H.}\ \bibnamefont {Park}}, \bibinfo {author} {\bibfnamefont {Z.}~\bibnamefont {Liu}}, \bibinfo {author} {\bibfnamefont {T.}~\bibnamefont {Yang}}, \bibinfo {author} {\bibfnamefont {Z.}~\bibnamefont {Zhang}}, \bibinfo {author} {\bibfnamefont {L.}~\bibnamefont {Shao}}, \bibinfo {author} {\bibfnamefont {X.}~\bibnamefont {Li}}, \bibinfo {author} {\bibfnamefont {T.}~\bibnamefont {Low}}, \bibinfo {author} {\bibfnamefont {Y.}~\bibnamefont {Shi}},\ and\ \bibinfo {author} {\bibfnamefont {X.}~\bibnamefont {Wang}},\ }\bibfield  {title} {\bibinfo
  {title} {Observation of chiral and slow plasmons in twisted bilayer graphene},\ }\href {https://doi.org/10.1038/s41586-022-04520-8} {\bibfield  {journal} {\bibinfo  {journal} {Nature}\ }\textbf {\bibinfo {volume} {605}},\ \bibinfo {pages} {63} (\bibinfo {year} {2022})}\BibitemShut {NoStop}%
\bibitem [{\citenamefont {Yin}\ \emph {et~al.}(2013)\citenamefont {Yin}, \citenamefont {Sch{\"a}ferling}, \citenamefont {Metzger},\ and\ \citenamefont {Giessen}}]{yin2013interpreting}%
  \BibitemOpen
  \bibfield  {author} {\bibinfo {author} {\bibfnamefont {X.}~\bibnamefont {Yin}}, \bibinfo {author} {\bibfnamefont {M.}~\bibnamefont {Sch{\"a}ferling}}, \bibinfo {author} {\bibfnamefont {B.}~\bibnamefont {Metzger}},\ and\ \bibinfo {author} {\bibfnamefont {H.}~\bibnamefont {Giessen}},\ }\bibfield  {title} {\bibinfo {title} {Interpreting chiral nanophotonic spectra: The plasmonic {Born--Kuhn} model},\ }\href {https://doi.org/10.1021/nl403705k} {\bibfield  {journal} {\bibinfo  {journal} {Nano Lett.}\ }\textbf {\bibinfo {volume} {13}},\ \bibinfo {pages} {6238} (\bibinfo {year} {2013})}\BibitemShut {NoStop}%
\bibitem [{\citenamefont {de~Vega}\ \emph {et~al.}(2020)\citenamefont {de~Vega}, \citenamefont {Cox}, \citenamefont {Sols},\ and\ \citenamefont {{Garc{\'\i}a de Abajo}}}]{devega2020strong}%
  \BibitemOpen
  \bibfield  {author} {\bibinfo {author} {\bibfnamefont {S.}~\bibnamefont {de~Vega}}, \bibinfo {author} {\bibfnamefont {J.~D.}\ \bibnamefont {Cox}}, \bibinfo {author} {\bibfnamefont {F.}~\bibnamefont {Sols}},\ and\ \bibinfo {author} {\bibfnamefont {F.~J.}\ \bibnamefont {{Garc{\'\i}a de Abajo}}},\ }\bibfield  {title} {\bibinfo {title} {Strong-field-driven dynamics and high-harmonic generation in interacting one-dimensional systems},\ }\href {https://doi.org/10.1103/PhysRevResearch.2.013313} {\bibfield  {journal} {\bibinfo  {journal} {Phys.\ Rev.\ Res.}\ }\textbf {\bibinfo {volume} {2}},\ \bibinfo {pages} {013313} (\bibinfo {year} {2020})}\BibitemShut {NoStop}%
\bibitem [{\citenamefont {Chen}\ \emph {et~al.}(2011)\citenamefont {Chen}, \citenamefont {Park}, \citenamefont {Boudouris}, \citenamefont {Horng}, \citenamefont {Geng}, \citenamefont {Girit}, \citenamefont {Zettl}, \citenamefont {Crommie}, \citenamefont {Segalman}, \citenamefont {Louie},\ and\ \citenamefont {Wang}}]{chen2011controlling}%
  \BibitemOpen
  \bibfield  {author} {\bibinfo {author} {\bibfnamefont {C.-F.}\ \bibnamefont {Chen}}, \bibinfo {author} {\bibfnamefont {C.-H.}\ \bibnamefont {Park}}, \bibinfo {author} {\bibfnamefont {B.~W.}\ \bibnamefont {Boudouris}}, \bibinfo {author} {\bibfnamefont {J.}~\bibnamefont {Horng}}, \bibinfo {author} {\bibfnamefont {B.}~\bibnamefont {Geng}}, \bibinfo {author} {\bibfnamefont {{\c{C}}.}~\bibnamefont {Girit}}, \bibinfo {author} {\bibfnamefont {A.}~\bibnamefont {Zettl}}, \bibinfo {author} {\bibfnamefont {M.~F.}\ \bibnamefont {Crommie}}, \bibinfo {author} {\bibfnamefont {R.~A.}\ \bibnamefont {Segalman}}, \bibinfo {author} {\bibfnamefont {S.~G.}\ \bibnamefont {Louie}},\ and\ \bibinfo {author} {\bibfnamefont {F.}~\bibnamefont {Wang}},\ }\bibfield  {title} {\bibinfo {title} {Controlling inelastic light scattering quantum pathways in graphene},\ }\href {https://doi.org/10.1038/nature09866} {\bibfield  {journal} {\bibinfo  {journal} {Nature}\ }\textbf {\bibinfo {volume} {471}},\ \bibinfo {pages} {617} (\bibinfo {year}
  {2011})}\BibitemShut {NoStop}%
\bibitem [{\citenamefont {Fei}\ \emph {et~al.}(2012)\citenamefont {Fei}, \citenamefont {Rodin}, \citenamefont {Andreev}, \citenamefont {Bao}, \citenamefont {McLeod}, \citenamefont {Wagner}, \citenamefont {Zhang}, \citenamefont {Zhao}, \citenamefont {Thiemens}, \citenamefont {Dominguez}, \citenamefont {Fogler}, \citenamefont {{Castro Neto}}, \citenamefont {Lau}, \citenamefont {Keilmann},\ and\ \citenamefont {Basov}}]{fei2012gatetuning}%
  \BibitemOpen
  \bibfield  {author} {\bibinfo {author} {\bibfnamefont {Z.}~\bibnamefont {Fei}}, \bibinfo {author} {\bibfnamefont {A.~S.}\ \bibnamefont {Rodin}}, \bibinfo {author} {\bibfnamefont {G.~O.}\ \bibnamefont {Andreev}}, \bibinfo {author} {\bibfnamefont {W.}~\bibnamefont {Bao}}, \bibinfo {author} {\bibfnamefont {A.~S.}\ \bibnamefont {McLeod}}, \bibinfo {author} {\bibfnamefont {M.}~\bibnamefont {Wagner}}, \bibinfo {author} {\bibfnamefont {L.~M.}\ \bibnamefont {Zhang}}, \bibinfo {author} {\bibfnamefont {Z.}~\bibnamefont {Zhao}}, \bibinfo {author} {\bibfnamefont {M.}~\bibnamefont {Thiemens}}, \bibinfo {author} {\bibfnamefont {G.}~\bibnamefont {Dominguez}}, \bibinfo {author} {\bibfnamefont {M.~M.}\ \bibnamefont {Fogler}}, \bibinfo {author} {\bibfnamefont {A.~H.}\ \bibnamefont {{Castro Neto}}}, \bibinfo {author} {\bibfnamefont {C.~N.}\ \bibnamefont {Lau}}, \bibinfo {author} {\bibfnamefont {F.}~\bibnamefont {Keilmann}},\ and\ \bibinfo {author} {\bibfnamefont {D.~N.}\ \bibnamefont {Basov}},\ }\bibfield  {title} {\bibinfo
  {title} {Gate-tuning of graphene plasmons revealed by infrared nano-imaging},\ }\href {https://doi.org/10.1038/nature11253} {\bibfield  {journal} {\bibinfo  {journal} {Nature}\ }\textbf {\bibinfo {volume} {487}},\ \bibinfo {pages} {82} (\bibinfo {year} {2012})}\BibitemShut {NoStop}%
\bibitem [{\citenamefont {Kim}\ \emph {et~al.}(2016)\citenamefont {Kim}, \citenamefont {S{\'a}nchez-Castillo}, \citenamefont {Ziegler}, \citenamefont {Ogawa}, \citenamefont {Noguez},\ and\ \citenamefont {Park}}]{kim2016chiral}%
  \BibitemOpen
  \bibfield  {author} {\bibinfo {author} {\bibfnamefont {C.-J.}\ \bibnamefont {Kim}}, \bibinfo {author} {\bibfnamefont {A.}~\bibnamefont {S{\'a}nchez-Castillo}}, \bibinfo {author} {\bibfnamefont {Z.}~\bibnamefont {Ziegler}}, \bibinfo {author} {\bibfnamefont {Y.}~\bibnamefont {Ogawa}}, \bibinfo {author} {\bibfnamefont {C.}~\bibnamefont {Noguez}},\ and\ \bibinfo {author} {\bibfnamefont {J.}~\bibnamefont {Park}},\ }\bibfield  {title} {\bibinfo {title} {Chiral atomically thin films},\ }\href {https://doi.org/10.1038/nnano.2016.3} {\bibfield  {journal} {\bibinfo  {journal} {Nat.\ Nanotechnol.}\ }\textbf {\bibinfo {volume} {11}},\ \bibinfo {pages} {520} (\bibinfo {year} {2016})}\BibitemShut {NoStop}%
\bibitem [{\citenamefont {Nam}\ and\ \citenamefont {Koshino}(2017)}]{nam2017lattice}%
  \BibitemOpen
  \bibfield  {author} {\bibinfo {author} {\bibfnamefont {N.~N.~T.}\ \bibnamefont {Nam}}\ and\ \bibinfo {author} {\bibfnamefont {M.}~\bibnamefont {Koshino}},\ }\bibfield  {title} {\bibinfo {title} {Lattice relaxation and energy band modulation in twisted bilayer graphene},\ }\href {https://doi.org/10.1103/PhysRevB.96.075311} {\bibfield  {journal} {\bibinfo  {journal} {Phys.\ Rev.\ B}\ }\textbf {\bibinfo {volume} {96}},\ \bibinfo {pages} {075311} (\bibinfo {year} {2017})}\BibitemShut {NoStop}%
\bibitem [{\citenamefont {{Castro Neto}}\ \emph {et~al.}(2009)\citenamefont {{Castro Neto}}, \citenamefont {Guinea}, \citenamefont {Peres}, \citenamefont {Novoselov},\ and\ \citenamefont {Geim}}]{castroneto2009electronic}%
  \BibitemOpen
  \bibfield  {author} {\bibinfo {author} {\bibfnamefont {A.~H.}\ \bibnamefont {{Castro Neto}}}, \bibinfo {author} {\bibfnamefont {F.}~\bibnamefont {Guinea}}, \bibinfo {author} {\bibfnamefont {N.~M.~R.}\ \bibnamefont {Peres}}, \bibinfo {author} {\bibfnamefont {K.~S.}\ \bibnamefont {Novoselov}},\ and\ \bibinfo {author} {\bibfnamefont {A.~K.}\ \bibnamefont {Geim}},\ }\bibfield  {title} {\bibinfo {title} {The electronic properties of graphene},\ }\href {https://doi.org/10.1103/RevModPhys.81.109} {\bibfield  {journal} {\bibinfo  {journal} {Rev.\ Mod.\ Phys.}\ }\textbf {\bibinfo {volume} {81}},\ \bibinfo {pages} {109} (\bibinfo {year} {2009})}\BibitemShut {NoStop}%
\bibitem [{\citenamefont {Mirzakhani}\ \emph {et~al.}(2025)\citenamefont {Mirzakhani}, \citenamefont {{\c{C}}etin}, \citenamefont {Ya{\u{g}}murcukarde{\c{s}}}, \citenamefont {Park}, \citenamefont {Peeters},\ and\ \citenamefont {{Da Costa}}}]{mirzakhani2025magnetism}%
  \BibitemOpen
  \bibfield  {author} {\bibinfo {author} {\bibfnamefont {M.}~\bibnamefont {Mirzakhani}}, \bibinfo {author} {\bibfnamefont {Z.}~\bibnamefont {{\c{C}}etin}}, \bibinfo {author} {\bibfnamefont {M.}~\bibnamefont {Ya{\u{g}}murcukarde{\c{s}}}}, \bibinfo {author} {\bibfnamefont {H.~C.}\ \bibnamefont {Park}}, \bibinfo {author} {\bibfnamefont {F.~M.}\ \bibnamefont {Peeters}},\ and\ \bibinfo {author} {\bibfnamefont {D.~R.}\ \bibnamefont {{Da Costa}}},\ }\bibfield  {title} {\bibinfo {title} {Magnetism in twisted triangular bilayer graphene quantum dots},\ }\href {https://doi.org/10.1103/PhysRevB.111.024417} {\bibfield  {journal} {\bibinfo  {journal} {Phys.\ Rev.\ B}\ }\textbf {\bibinfo {volume} {111}},\ \bibinfo {pages} {024417} (\bibinfo {year} {2025})}\BibitemShut {NoStop}%
\bibitem [{\citenamefont {Yoshizawa}\ \emph {et~al.}(2000)\citenamefont {Yoshizawa}, \citenamefont {Yumura}, \citenamefont {Yamabe},\ and\ \citenamefont {Bandow}}]{yoshizawa2000orbital}%
  \BibitemOpen
  \bibfield  {author} {\bibinfo {author} {\bibfnamefont {K.}~\bibnamefont {Yoshizawa}}, \bibinfo {author} {\bibfnamefont {T.}~\bibnamefont {Yumura}}, \bibinfo {author} {\bibfnamefont {T.}~\bibnamefont {Yamabe}},\ and\ \bibinfo {author} {\bibfnamefont {S.}~\bibnamefont {Bandow}},\ }\bibfield  {title} {\bibinfo {title} {The role of orbital interactions in determining the interlayer spacing in graphite slabs},\ }\href {https://doi.org/10.1021/ja994457o} {\bibfield  {journal} {\bibinfo  {journal} {J.\ Am.\ Chem.\ Soc.}\ }\textbf {\bibinfo {volume} {122}},\ \bibinfo {pages} {11871} (\bibinfo {year} {2000})}\BibitemShut {NoStop}%
\bibitem [{\citenamefont {Cox}\ \emph {et~al.}(2017)\citenamefont {Cox}, \citenamefont {Marini},\ and\ \citenamefont {{Garc{\'\i}a de Abajo}}}]{cox2017plasmonassisted}%
  \BibitemOpen
  \bibfield  {author} {\bibinfo {author} {\bibfnamefont {J.~D.}\ \bibnamefont {Cox}}, \bibinfo {author} {\bibfnamefont {A.}~\bibnamefont {Marini}},\ and\ \bibinfo {author} {\bibfnamefont {F.~J.}\ \bibnamefont {{Garc{\'\i}a de Abajo}}},\ }\bibfield  {title} {\bibinfo {title} {Plasmon-assisted high-harmonic generation in graphene},\ }\href {https://doi.org/10.1038/ncomms14380} {\bibfield  {journal} {\bibinfo  {journal} {Nat.\ Commun.}\ }\textbf {\bibinfo {volume} {8}},\ \bibinfo {pages} {14380} (\bibinfo {year} {2017})}\BibitemShut {NoStop}%
\bibitem [{\citenamefont {Cox}\ and\ \citenamefont {{Garc{\'\i}a de Abajo}}(2019)}]{cox2019singleplasmon}%
  \BibitemOpen
  \bibfield  {author} {\bibinfo {author} {\bibfnamefont {J.~D.}\ \bibnamefont {Cox}}\ and\ \bibinfo {author} {\bibfnamefont {F.~J.}\ \bibnamefont {{Garc{\'\i}a de Abajo}}},\ }\bibfield  {title} {\bibinfo {title} {Single-plasmon thermo-optical switching in graphene},\ }\href {https://doi.org/10.1021/acs.nanolett.9b00879} {\bibfield  {journal} {\bibinfo  {journal} {Nano Lett.}\ }\textbf {\bibinfo {volume} {19}},\ \bibinfo {pages} {3743} (\bibinfo {year} {2019})}\BibitemShut {NoStop}%
\bibitem [{\citenamefont {Jelver}\ and\ \citenamefont {Cox}(2024)}]{jelver2024nonlinear}%
  \BibitemOpen
  \bibfield  {author} {\bibinfo {author} {\bibfnamefont {L.}~\bibnamefont {Jelver}}\ and\ \bibinfo {author} {\bibfnamefont {J.~D.}\ \bibnamefont {Cox}},\ }\bibfield  {title} {\bibinfo {title} {Nonlinear thermoplasmonics in graphene nanostructures},\ }\href {https://doi.org/10.1021/acs.nanolett.4c04048} {\bibfield  {journal} {\bibinfo  {journal} {Nano Lett.}\ }\textbf {\bibinfo {volume} {24}},\ \bibinfo {pages} {13775} (\bibinfo {year} {2024})}\BibitemShut {NoStop}%
\end{thebibliography}

%

\appendix

\section{Supplementary Figures}

\begin{figure*}[hbt]
    \centering
    \includegraphics[width=0.9\linewidth]{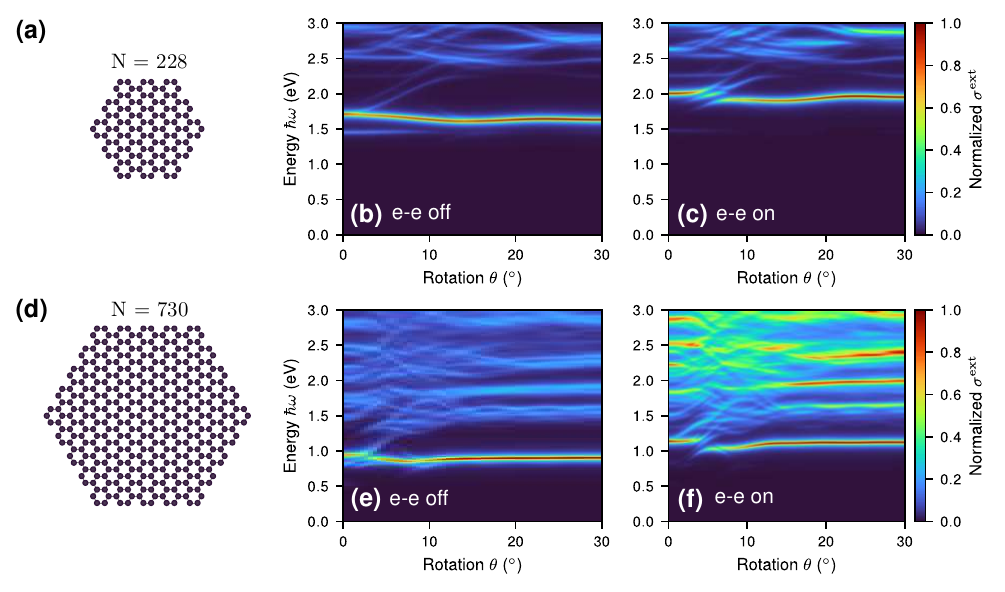}
    \caption{\textbf{Effect of electron--electron interactions on the linear optical response of AA-stacked hexagonal TBG nanoislands with armchair edge terminations.} \textbf{(a)} Atomic positions of the smaller hexagonal TBG nanoisland ($N=228$). External electric field $E^{\text{ext}}$ polarization indicated. \textbf{(b,c)} Linear optical response, characterized by the normalized extinction cross section $\sigma^{\text{ext}}$ of the undoped small TBG bilayer with electron--electron (e-e) interactions neglected and included. \textbf{(d)} Atomic positions of the larger hexagonal TBG nanoisland ($N=730$). \textbf{(e,f)} Same as (b,c), but for the larger bilayer system.}
    \label{figS1}
\end{figure*}

\begin{figure*}[hbt]
    \centering
    \includegraphics[width=0.8\linewidth]{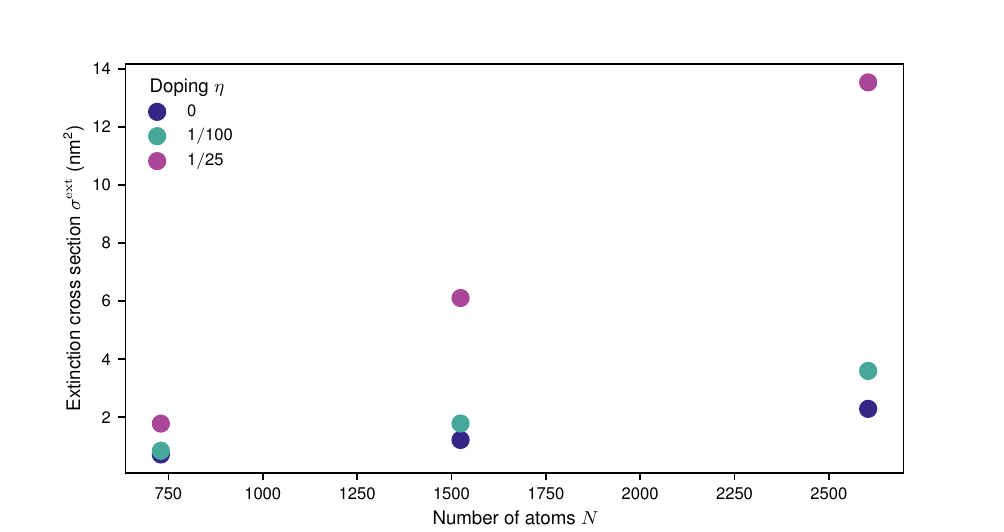}
    \caption{\textbf{Size and doping dependence of the peak extinction of hexagonal TBG nanoislands.} Maximum extinction cross section as a function of size $N$ and doping level $\eta$ of the AA-stacked hexagonal TBG nanoislands shown in Fig.\ 2 of the main text, with armchair edge termination.}
    \label{figS2}
\end{figure*}

\begin{figure*}[hbt]
    \centering
    \includegraphics[width=0.9\linewidth]{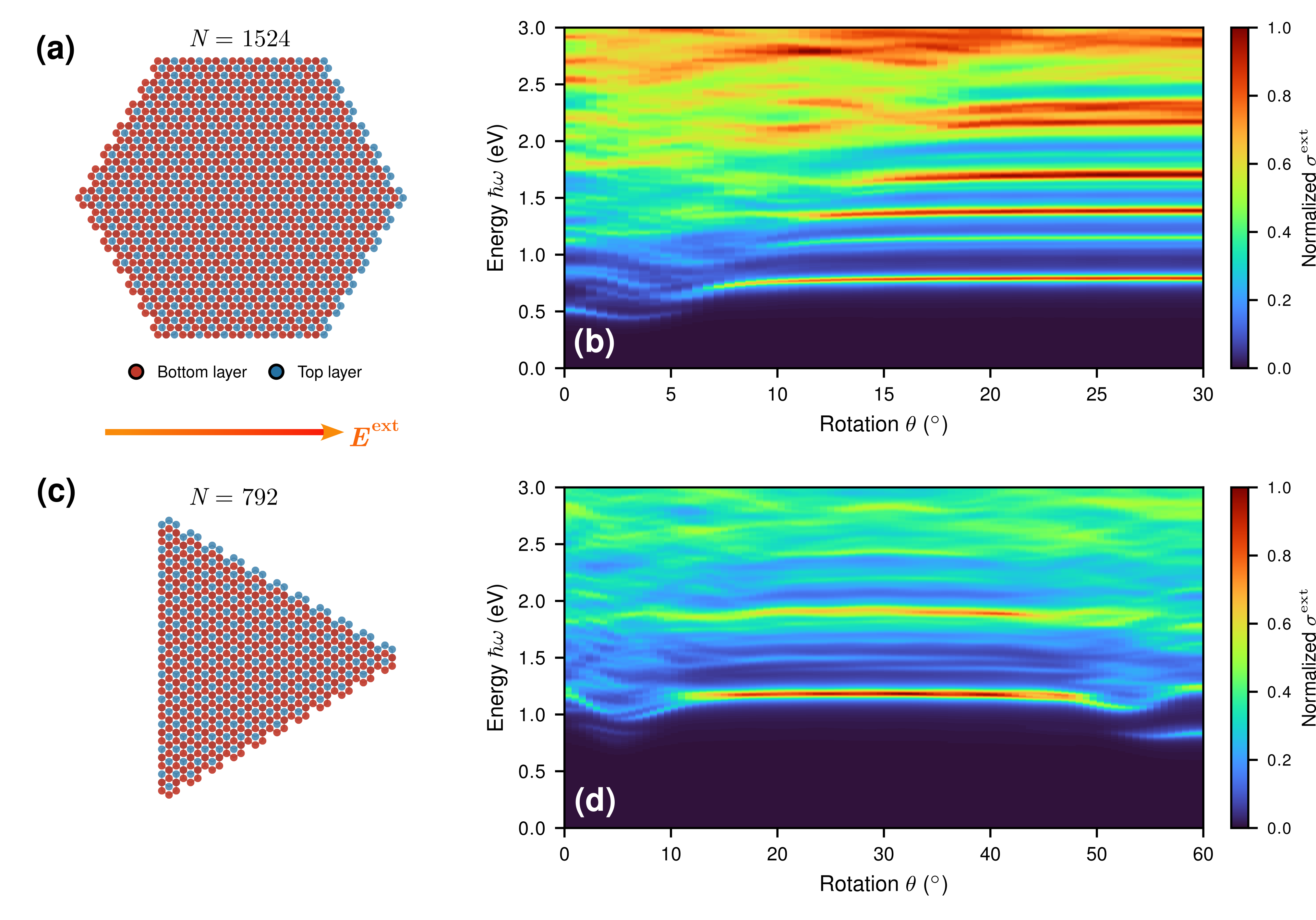}
    \caption{\textbf{Bernal (AB) stacking and linear optical response of hexagonal and triangular TBG nanoislands with armchair edge termination.} \textbf{(a)} Atomic positions of the hexagonal TBG nanoisland ($\theta=0\mydeg$) with $N=1524$. External electric field $E^{\text{ext}}$ polarization indicated. \textbf{(b)} Extinction cross section normalized to the maximum peak, as a function of energy and twist angle for the undoped hexagonal TBG. \textbf{(c)} Triangular TBG nanoisland ($\theta=0\mydeg$) with $N=792$. \textbf{(d)} Same as (b) but for the triangular TBG.}
    \label{figS3}
\end{figure*}

\begin{figure*}[hbt]
    \centering
    \includegraphics[width=0.9\linewidth]{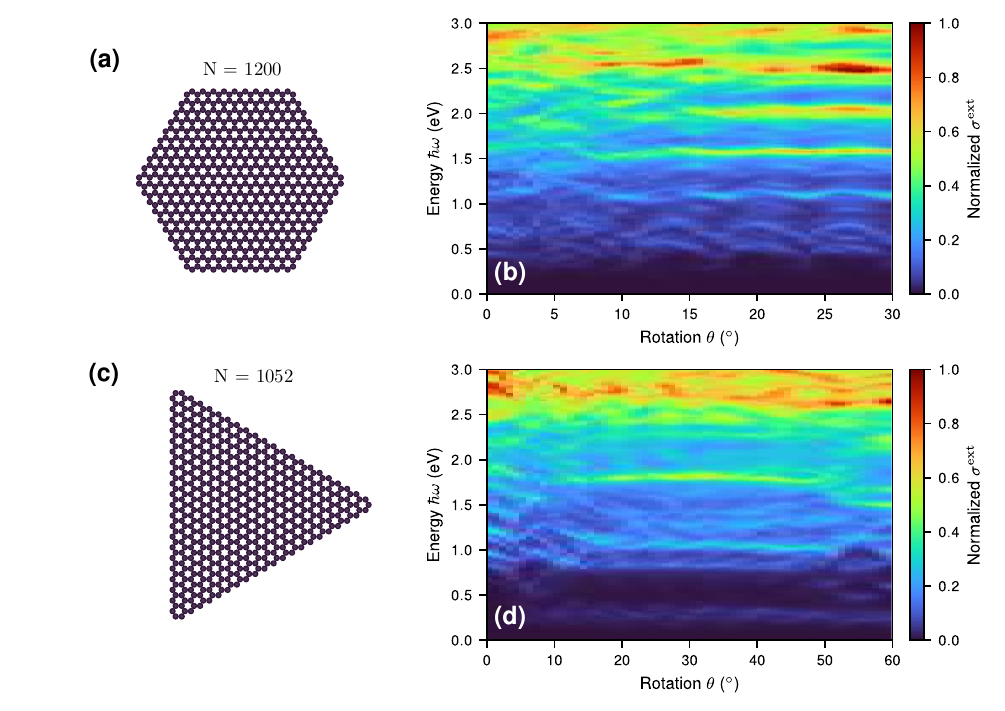}
    \caption{\textbf{Linear optical response for AA-stacked hexagonal and triangular TBG nanoislands, with zigzag edge-terminations.} \textbf{(a)} Atomic positions of the hexagonal TBG nanoisland ($\theta=0\mydeg$) with $N=1200$. External electric field $E^{\text{ext}}$ polarization indicated. \textbf{(b)} Extinction cross section normalized to the maximum peak, as a function of energy and twist angle for the undoped hexagonal TBG. \textbf{(c)} Triangular TBG nanoisland ($\theta=0\mydeg$) with $N=1052$. \textbf{(d)} Same as (b) but for the triangular TBG.}
    \label{figS4}
\end{figure*}

We present additional simulations that support the main-text analysis of twisted bilayer graphene (TBG) nanoislands. We first show that the abrupt reorganization of the extinction spectrum at the critical moir\'e angle $\theta_{\rm m}$ disappears when electron--electron interactions are neglected, confirming its collective origin (Fig.~\ref{figS1}). We then quantify how the peak extinction cross section of AA-stacked hexagonal islands grows with island size and doping level (Fig.~\ref{figS2}). Finally, we demonstrate that the twist-dependent spectral features reported in the main text persist for Bernal (AB) stacking (Fig.~\ref{figS3}), and examine how zigzag edge terminations modify the energies and intensities of the moir\'e-commensurate resonances (Fig.~\ref{figS4}), for both hexagonal and triangular islands.

\end{document}